\documentclass[sn-mathphys,Numbered]{sn-jnl}

\usepackage{graphicx}         
\usepackage{multirow}         
\usepackage{amsmath, amssymb, amsfonts}   
\usepackage{amsthm}            
\usepackage{mathrsfs}         
\usepackage{xcolor}           
\usepackage{textcomp}         
\usepackage{manyfoot}         
\usepackage{caption}          
\usepackage{longtable}        
\usepackage{siunitx}          
\usepackage{algorithm}        
\usepackage{algorithmicx}     
\usepackage{algpseudocode}    
\usepackage{listings}         
\usepackage{nomencl}          
\usepackage{lineno}           
\usepackage{txfonts}          
\usepackage{appendix}         
\usepackage{tabularx}         

\usepackage{nomencl}
\makenomenclature

\usepackage{booktabs}
\DeclareSIUnit{\kgw}{kgw}     
\DeclareSIUnit{\kgo}{kgo}     

\begin{document}
	
	\title{Estimation of oil recovery due to wettability changes in carbonate reservoirs}
	
	\author*[1]{\fnm{A.C.} \sur{Alvarez}}\email{amaury@ic.ufrj.br}
	\equalcont{These authors contributed equally to this work.}
	
	\author[2]{\fnm{J.} \sur{Bruining}}\email{J.Bruining@tudelft.nl}
	\equalcont{These authors contributed equally to this work.}
	
	\author[3]{\fnm{D.} \sur{Marchesin}}\email{marchesi@impa.br}
	\equalcont{These authors contributed equally to this work.}
	
	\affil*[1]{\orgdiv{Instituto de Computação}, \orgname{Universidade Federal do Rio de Janeiro}, \orgaddress{\street{Av. Athos da Silveira Ramos, 274}, \city{Rio de Janeiro}, \postcode{21941-590}, \state{Rio de Janeiro}, \country{Brasil}}}
				
	\affil[2]{\orgdiv{ Civil Engineering and Geosciences}, \orgname{TU Delft}, \orgaddress{\street{ Stevinweg 1}, \city{Delft}, \postcode{2628 CE}, \state{Delft}, \country{The Netherlands}}}
		
	\affil[3]{\orgdiv{Lab. Fluid Dynamics}, \orgname{IMPA}, \orgaddress{\street{Estrada Dona Castorina, 110}, \city{Rio de Janeiro}, \postcode{22460-320}, \state{Rio de Janeiro}, \country{Brasil}}}

\abstract{
Low salinity waterflooding (LSWF) enhances oil recovery at low cost in carbonate reservoirs, but its effectiveness requires precise control of injected water chemistry and interaction with reservoir minerals. This study develops an integrated reactive transport model coupling geochemical surface complexation modeling (SCM) with multiphase compositional dynamics to quantify wettability alteration during LSWF. The framework combines PHREEQC-based equilibrium calculations of the Total Bond Product (TBP)—a wettability indicator derived from oil–calcite ionic bridging—with Corey-type relative permeability interpolation, resolved via COMSOL Multiphysics. Core flooding simulations, compared with experimental data from calcite systems at \SI{100}{\celsius} and \SI{220}{\bar}, reveal that magnesium ([Mg$^{2+}$]) and sulfate ([SO$_4^{2-}$]) concentrations modulate TBP, reducing oil–rock adhesion under controlled low-salinity conditions. Parametric analysis demonstrates that acidic crude oils (TAN higher than \SI{1}{\milli\gram\ KOH\per\gram}) exhibit TBP values approximately \(\SI{2.5}{times}\) higher than sweet crudes, due to carboxylate–calcite bridging, while pH elevation (higher than 7.5) amplifies wettability shifts by promoting deprotonated -{COO$^{-}$} interactions. The model further identifies synergistic effects between ([Mg$^{2+}$]) (ranging from 50 to 200 mmol/kgw) and ([SO$_4^{2-}$]) (higher than 500 mmol/kgw), which reduce ({Ca$^{2+}$})-mediated oil adhesion through competitive mineral surface binding. By correlating TBP with fractional flow dynamics, this framework could support the optimization of injection strategies in carbonate reservoirs, suggesting that ion-specific adjustments are more effective than bulk salinity reduction.			
}



\maketitle

\section*{Nomenclature}

\begin{longtable}{@{}llc@{}}
	\toprule
	Symbol & Description & Unit \\
	\midrule
	\endfirsthead
	
	\toprule
	Symbol & Description & Unit \\
	\midrule
	\endhead
	
	\midrule
	\endfoot
	
	\bottomrule
	\endlastfoot
	
	\toprule
	Symbol & Description & Unit \\
	\midrule
	\endhead
	
	\midrule
	\endfoot
	
	\bottomrule
	\endlastfoot
	
	\(N_{s}\) & Number of chemical species & Dimensionless \\
	\(n_{s}\) & Number of surface species & Dimensionless \\
	\(N_{r}\) & Number of chemical reactions & Dimensionless \\
	\(n_{c}\) & Number of constraints & Dimensionless \\
	\(n_{f}\) & Degrees of freedom & Dimensionless \\
	\(S_w\) & Water saturation & Dimensionless \\
	\(S_o\) & Oil saturation & Dimensionless \\
	\(k_{rw}\) & Relative permeability of water & Dimensionless \\
	\(k_{ro}\) & Relative permeability of oil & Dimensionless \\
	\(n_w\) & Corey exponent for water & Dimensionless \\
	\(n_o\) & Corey exponent for oil & Dimensionless \\
	\(P_c\) & Capillary pressure & \si{\pascal} \\
	TBP & Total Bond Product & Dimensionless \\
	\(u\) & Darcy velocity & \si{\meter\per\second} \\
	\(\mu_w\) & Water viscosity & \si{\pascal\second} \\
	\(\mu_o\) & Oil viscosity & \si{\pascal\second} \\
	\([\text{Cl}^-]\) & Chloride ion concentration & \si{\milli\mol\per\kilogram} \\
	\([\text{Na}^+]\) & Sodium ion concentration & \si{\milli\mol\per\kilogram} \\
	\([\text{Mg}^{2+}]\) & Magnesium ion concentration & \si{\milli\mol\per\kilogram} \\
	\([\text{Ca}^{2+}]\) & Calcium ion concentration & \si{\milli\mol\per\kilogram} \\
	\([\text{SO}_4^{2-}]\) & Sulfate ion concentration & \si{\milli\mol\per\kilogram} \\
	pH & Hydrogen ion activity & Dimensionless \\
	\(\varphi\) & Porosity & Dimensionless \\
	\(\theta\) & Interpolation parameter & Dimensionless \\
	\(T\) & Temperature & \si{\celsius} \\
	\(P\) & Pressure & \si{\bar} \\
	TAN & Total Acid Number & \si{\milli\gram\ KOH\per\gram} \\
	TBN & Total Base Number & \si{\milli\gram\ KOH\per\gram} \\
	\(a_{\text{oil}}\) & Oil specific surface area & \si{\square\meter\per\gram} \\
	\(\rho_{w}\) & Molar density of aqueous phase & \si{\milli\mol\per\kgw} \\
	\(\rho_{o}\) & Molar density of oleic phase & \si{\milli\mol\per\kgo} \\
	\(x_{wi}\) & Molar fraction of component \(i\) in aqueous phase & Dimensionless \\
	\(x_{oj}\) & Molar fraction of component \(j\) in oleic phase & Dimensionless \\
	\(\rho_{rj}\) &molar fraction of the component j in the solid phase & Dimensionless \\
	\(\rho_{wi}\) & Concentration of component \(i\) in water phase (\(\rho_w x_{wi}\)) & \si{\milli\mol\per\kgw} \\
	\(\rho_{oi}\) & Concentration of component \(i\) in oleic phase (\(\rho_o x_{oj}\)) & \si{\milli\mol\per\kgo} \\
	\(f_w(S_w)\) & Fractional flow function for water & Dimensionless \\
	\(f_o(S_o)\) & Fractional flow function for oil & Dimensionless \\
	\(c_{a,i}\) & Concentration of aqueous species \(i\) & \si{\milli\mol\per\kgw} \\
	\(c_{o,i}\) & Concentration of oleic species \(i\) & \si{\milli\mol\per\kgo} \\
	\(c_{r,i}\) & Concentration of solid species \(i\) & \si{\milli\mol\per\kgw} \\
	\(\text{Oil}_{x,Y}\) & Oil surface complex (site \(x\), species \(Y\)) & \si{\milli\mol\per\kgw} \\
	\(\text{Cal}_{x,Y}\) & Calcite surface complex (site \(x\), species \(Y\)) & \si{\milli\mol\per\kgw} \\
	\(\mu\) & Ionic strength of the aqueous phase & Dimensionless \\
	\(\beta_{\mathrm{Ca}}\), \(\beta_{\mathrm{Mg}}\) & Precipitated amounts of calcium and magnesium & \si{\milli\mol\per\kgw} \\
	\(C_{a,X}\) & Total concentration of component \(X\) in aqueous phase & \si{\milli\mol\per\kgw} \\
	\(GA\) & Total mass of aqueous phase & \si{\kgw} \\
	\(GO\) & Total mass of oleic phase & \si{\kgo} \\
	\midrule
	\multicolumn{3}{l}{\textbf{Key species indices (\(i\)):}} \\
	\multicolumn{3}{l}{Aqueous (\(c_{a,i}\)):} \\
	\multicolumn{3}{l}{\(\text{CO}_2, \text{CO}_3^{2-}, \text{HCO}_3^{-}, \text{CaHCO}_3^{+}, \text{CaCO}_3, \text{NaCO}_3^{-}, \text{NaHCO}_3, \text{H}_2\text{O}, \text{H}^+, \text{OH}^-,\)} \\
	\multicolumn{3}{l}{\(\text{CaOH}^+, \text{Ca}^{2+}, \text{Mg}^{2+}, \text{MgHCO}_3^{+}, \text{MgCO}_3^0, \text{MgOH}^+, \text{Cl}^-, \text{Na}^+,\text{SO}_4^{2-},\)} \\
	\multicolumn{3}{l}{\(\text{CaSO}_4, \text{MgSO}_4, \text{NaSO}_4^{-}, \text{H}_2\text{SO}_4\)} \\
	\multicolumn{3}{l}{Oleic (\(c_{o,i}\)): \(\text{A}\) (alkane)} \\
	\multicolumn{3}{l}{Solid (\(c_{r,i}\)): \(\text{CaCO}_3\) (partitioned), \(\text{CaSO}_4\)} \\
	\multicolumn{3}{l}{Sorbed (\(\text{Oil}_{x,Y}\), \(\text{Cal}_{x,Y}\)):} \\
	\multicolumn{3}{l}{\(\text{Oil}_{s,\text{NH}^+}, \text{Oil}_{w,\text{COOH}}, \text{Cal}_{s,\text{OH}}, \text{Cal}_{w,\text{CO}_3\text{H}}, \text{Oil}_{s,\text{N}}, \text{Oil}_{w,\text{COO}^-},\)} \\
	\multicolumn{3}{l}{\(\text{Oil}_{w,\text{COOCa}^+}, \text{Oil}_{w,\text{COOMg}^+}, \text{Cal}_{s,\text{OH}_2^+}, \text{Cal}_{s,\text{CO}_3^-}, \text{Cal}_{w,\text{CO}_3^-},\)} \\
	\multicolumn{3}{l}{\(\text{Cal}_{w,\text{CO}_3\text{Ca}^+}, \text{Cal}_{w,\text{CO}_3\text{Mg}^+}, \text{Cal}_{s,\text{SO}_4^-}, \text{Oil}_{s,\text{NH}_2\text{SO}_4^-}\)} \\
	\midrule
	\midrule
	\multicolumn{3}{l}{\textbf{Key species indices (\(i\)):}} \\
	\midrule
	\multicolumn{3}{l}{\textbf{Aqueous (\(C_{a,i}\)):}} \\
	\(C_{a,C(4)}\) & Total inorganic carbon (e.g., \(CO_3^{2-}\), \(HCO_3^{-}\)) & \si{\milli\mol\per\kgw} \\
	\(C_{a,H(1)}\) & Total hydrogen ion activity  & Dimensionless \\
	\(C_{a,O(-2)}\) & Total oxygen in water (\(H_2O\), \(OH^{-}\)) & \si{\milli\mol\per\kgw} \\
	\(C_{a,Ca^{2+}}\) & Total calcium ion concentration & \si{\milli\mol\per\kgw} \\
	\(C_{a,Mg^{2+}}\) & Total magnesium ion concentration & \si{\milli\mol\per\kgw} \\
	\(C_{a,Cl^{-}}\) & Total chloride ion concentration & \si{\milli\mol\per\kgw} \\
	\(C_{a,Na^{+}}\) & Total sodium ion concentration & \si{\milli\mol\per\kgw} \\
	\midrule
	\multicolumn{3}{l}{\textbf{Oleic (\(C_{o,i}\)):}} \\
	\(C_{o,CO_2}\) & Dissolved carbon dioxide in oleic phase & \si{\milli\mol\per\kgo} \\
	\(C_{o,C(-4)}\) & Organic carbon (e.g., alkane \(A\)) & \si{\milli\mol\per\kgo} \\
	\midrule
	\multicolumn{3}{l}{\textbf{Solid/Sorbed (\(c_{r,i}\), \(Cal_{x,Y}\)):}} \\
	\(C_{s,CaCO_3}\) & Solid calcium carbonate & \si{\milli\mol\per\kgw} \\
	\(Cal_{s,CO_3^-}\) & Carbonate ions in calcite structure & \si{\milli\mol\per\kgw} \\
	\(Cal_{s,OH_2^+}\) & Hydroxylated calcite surface (\(>CaOH_2^+\)) & \si{\milli\mol\per\kgw} \\
	\(Cal_{s,OH}\) & Neutral hydroxylated calcite site (\(>CaOH\)) & \si{\milli\mol\per\kgw} \\
	\(Cal_{w,CO_3H}\) & Surface-bound bicarbonate (\(>CaCO_3H\)) & \si{\milli\mol\per\kgw} \\
	\(Cal_{w,CO_3Mg^+}\) & Magnesium-carbonate surface complex & \si{\milli\mol\per\kgw} \\
	\(Cal_{w,CO_3Ca^+}\) & Calcium-carbonate surface complex & \si{\milli\mol\per\kgw} \\
	\midrule
\end{longtable}

\section{Introduction}
\label{intro}
		
Low-salinity waterflooding (LSWF) consists of the injection of water with lower salinity and different ionic composition from the connate water to enhance oil recovery (EOR) (\cite{dong2011experimental,ayirala2015state,sharma2018experimental}). The methodology of LSWF includes modifying the salinity and ionic composition of the water used for waterflooding. This method is used as a secondary recovery method, which occurs after the reservoir natural energy has been depleted. The utilization of LSWF is cost-effective and environmentally viable, with recent advancements highlighting its potential to reduce interfacial tension and mobilize residual oil through ion-specific interactions \cite{alqattan2023dynamic,lee2024field}.

	The
modification of salinity and particular ions is often
	known as designed, dynamic or smart water (\cite{Zhang2007,snosy2022comprehensive}).
	Many processes have been proposed to explain
	the EOR achieved during LSWF in  carbonate reservoirs.
	These processes can be  divided into two main groups: fluid–fluid
	interactions (\cite{lsanda37})) and rock-fluid interactions (\cite{barte42}). The rock-fluid interactions are studied mainly through the wettability measurements. Existing wettability alteration mechanisms include multi-component ionic exchange,
	expansion of the electrical double layer, rock dissolution, fines migration, interfacial tension decrease, pH increase and formation of microdispersions.

	LSWF with smart water is the basis of the fact that modified ion water composition produces a loss of the equilibrium in the initial oil-rock-brine system.
	This loss produces changes in the initial wettability conditions (\cite{arain24}). Several studies present relevant advances to explain how changes in water ions composition produces an increase in oil recovery during waterflooding processes (see examples in \cite{gba22}). However, this oil recovery technique still requires further clarification to achieve optimal results.

	 This study explores, through numerical simulations, the influence of the main ion concentrations that affect the wettability, i.e., magnesium, calcium, sulfate, sodium, and chloride ions (\cite{jerauld2008modeling,appelo2005geochemistry,austad2015low,yousef2012improved,Chakravarty2015}). Here we utilize the multiphase compositional modeling technique like in \cite{alvarez2018analytical1,alvarez2018analytical,sanaei2019mechanistic}. We assess factors like pH, type of oil and the concentration of key ions, to quantify how the complexes alter wettability (\cite{al2015geochemical}).

	We develop a model that integrates compositional phase and geochemical modeling through a system of conservation laws, as described in \cite{alvarez2024modeling}. First, we define the model coefficients, accounting for ion transport. Gibbs' phase rule is used to determine the number of the degrees of freedom. The model coefficients are then expressed as functions of selected free master species. (\cite{alvarez2018analytical1}).  We utilize the PHREEQC program (see details in \cite{parkhurst1999user}) to estimate the concentration of the master species.
	
	In  \cite{austad2015low} 
	 an experimental study of the oil recovery after carbonate water injection in a rock formed mainly by calcite and some  anhydrite is presented. The experimental data from \cite{austad2015low} are used to assess the model.
  	
  	 In \cite{mehdiyev2022surface} is propose TBP as a proxy for wettability estimation via interfacial bond energies. Our contribution, however, 
  	is to use it also as an interpolation parameter between the regimes of high and low salinity. The TBP parameter interpolates capillary pressure and relative permeability.
 
  	The Total Bond Product emerges as a critical wettability indicator by quantifying the cumulative strength of ionic bridges (e.g., Ca$^{2+}$–carboxylate, Mg$^{2+}$–sulfate) at oil-calcite interfaces. Unlike conventional metrics like contact angle, TBP directly correlates surface complexation thermodynamics with macroscopic displacement efficiency, enabling predictive optimization of injection strategies. Its derivation from equilibrium species calculated with PHREEQC (\cite{mehdiyev2022surface,Bonto2019}) provides a mechanistic link between brine chemistry and oil recovery.
  	
  	 The SCM is a chemical equilibrium model adept at simulating the  exchanges among water, oil, brine, and rock \cite{lutzenkirchen2006surface,elakneswaran2017surface}. This model facilitates the characterization of surface adsorption phenomena \cite{marmier1997surface,sanaei2019investigation} and the determination of mineral wettability under reservoir conditions \cite{bordeaux2021improvements,mehdiyev2022surface}. We calculate  We implement the Surface Complexation Model (SCM) using the PHREEQC program, with a focus on calculating the Total Bond Product (TBP) as a key wettability indicator, following the methodology detailed in \cite{Bonto2019}.
	
	Numerous studies have explored the correlation between salt concentration, oil type and water flooding processes, as evidenced by the works of \cite{jerauld2008modeling,omekeh2012modeling,al2015novel}. 
  	We base our study on \cite{jerauld2008modeling,alvarez2018analytical,alvarez2018analytical} and use Corey permeability function (\cite{bruining2021upscaling}).

	The methodology outlined in this study  provides a unified framework for integrating geochemistry and compositional modeling. Furthermore, this approach facilitates the exploration of several scenarios. By using Gibbs rule, we streamline the mathematical complexity associated with physical constraints and relevant parameters for the geochemical simulations.
	
	Here, we extend the approach in \cite{alvarez2018analytical,alvarez2018analytical1,alvarez2024modeling}, which focuses solely on chloride and sodium ions.  Indeed, our method allows to incorporate other ions.

    This study pursues two interconnected objectives: (1) to quantify the sensitivity of relative permeability and oil recovery to key divalent ions (Mg\textsuperscript{2+}, SO\textsubscript{4}\textsuperscript{2-}, Ca\textsuperscript{2+}) in carbonate reservoirs, and (2) to elucidate the mechanistic role of oil-calcite surface complexes in wettability alteration. Using the Total Bond Product (TBP) as a thermodynamic metric, we systematically evaluate how brine chemistry, crude oil composition (acidity via TAN, basicity via TBN), and reservoir conditions (\SI{100}{\celsius}, \SI{220}{\bar}) collectively govern adhesion dynamics during low-salinity waterflooding.
    
	This paper is structured as follows.
	
	Section \ref{che1} develops the physical-chemical model for carbonate reservoirs, including aqueous/sorbed species interactions and Gibbs phase rule analysis. Section \ref{geo} formulates the governing equations for multiphase flow, ion transport, and mass conservation. Section \ref{geo2} details the geochemical modeling framework using PHREEQC, incorporating TAN/TBN correlations and surface complexation reactions. 
	Section \ref{frac} introduces the fractional flow model with Corey-type permeability interpolation, linking it to TBP. In the Section \ref{int} 
	is presented a summary of integration processes of Geochemical and Multiphase modeling.
	Section \ref{wet} systematically evaluates wettability alteration mechanisms through parametric studies of pH, Mg$^{2+}$, SO$_4^{2-}$, and Ca$^{2+}$-SO$_4^{2-}$ synergies. Section \ref{sec:TBP_theta_analysis} quantifies ionic bridging effects via TBP-driven analysis under variable salinity regimes. In the Section \ref{sec:TBP_validation}
	is described a comparison of TBP-based and experimental wettability metrics.
	Section \ref{sec:numerical} assesses the integrated geochemical flow model against coreflood experiments using COMSOL simulations. Section \ref{sec:salinity_equilibrium} reconciles injected vs.\ equilibrium salinity through thermodynamic activity principles. Sections \ref{con} present conclusions.
	Finally, in the appendix are presented coefficient derivations, tables and supporting figures.

\section{Physical-Chemical  Model}
\label{che1}
We extend our analysis to model aqueous and sorbed species in a carbonate reservoir system. The aqueous phase includes ions (\(\text{H}^+\), \(\text{OH}^-\), \(\text{CO}_3^{2-}\), \(\text{HCO}_3^-\), \(\text{Cl}^-\), \(\text{Mg}^{2+}\), \(\text{Ca}^{2+}\), \(\text{SO}_4^{2-}\)) and water (\(\text{H}_2\text{O}\)), while sorbed species encompass oil-calcite complexes (e.g., \(\text{oil}_{s,\text{NH}^+}\), \(\text{oil}_{w,\text{COOH}}\), \(\text{Cal}_{s,\text{OH}}\), \(\text{Cal}_{w,\text{CO}_3\text{H}}\)) \cite{Brady2012}. These interactions form the basis of  the \textit{Surface Complexes-Chloride Ionic Carbon Dioxide-Oil-Water (SC-CLICDOW)} model, which integrates equilibrium thermodynamics, ion transport, and wettability dynamics \cite{helfferich1989theory}. To evaluate this framework, we conduct core flooding experiments with pH-matched, low-salinity carbonated water, targeting enhanced oil recovery (EOR) through salinity reduction and divalent cation (\(\text{Mg}^{2+}\), \(\text{Ca}^{2+}\), \(\text{SO}_4^{2-}\)) modulation.  

The reservoir is modeled as a 1D porous medium saturated with oleic and aqueous phases. Initial and injected fluids contain NaCl, CO\(_2\), and key ions. Carbon dioxide partitions between oleic and aqueous phases and decane remains only in the oleic phase. Rapid equilibrium assumptions apply to aqueous-oleic CO\(_2\) exchange and geochemical reactions, simplifying the ion transport analysis. Darcy's law governs incompressible flow at \(T = \SI{100}{\celsius}\) and \(P = \SI{220}{\bar}\), suppressing gas phase formation. We neglect salinity-dependent viscosity \cite{austad2015low}. 

	The assumption of salinity-independent viscosity is justified by two main factors. First, at reservoir conditions, particularly at high temperatures around \SI{100}{\celsius}, the viscosity contrast between injected low-salinity brines and formation high-salinity brines becomes negligible (see .e.g.,\cite{sharma2020enhanced}). Second, in carbonate reservoirs, oil recovery is primarily governed by geochemical wettability alterations caused by ionic interactions, such as the exchange between \(\text{Ca}^{2+}\) and \(\text{Mg}^{2+}\), rather than by fractional flow changes resulting from minor viscosity contrasts \cite{mahani2015}.

The analysis suggests that carbonated low-salinity waterflooding increases dissolved $\text{CO}_2$ levels, contributing to calcite dissolution and higher aqueous $\text{Ca}^{2+}$, $\text{Mg}^{2+}$, and $\text{SO}_4^{2-}$ concentrations.
This ionic shift correlates with reduced oil-rock adhesion through changes in Total Bond Product (TBP)-associated wettability, assessed using high/low-salinity relative permeability curves. The synergy between CO\(_2\) solubility and controlled divalent cation availability amplifies oil mobilization, consistently with EOR mechanisms reported in \cite{lake2014}.  

\subsection{Chemical Equilibrium Analysis: Gibbs rule}
Utilizing the methodologies outlined in \cite{parkhurst1999user,appelo2005geochemistry}, we employed PHREEQC to simulate the equilibrium of water, solid calcium carbonate (\( \text{CaCO}_{3(\text{solid})} \)), sodium chloride (\( \text{NaCl} \)), and sulfate species. Our analysis identifies {24 distinct chemical species} (\( N_s = 24 \)) in the system:
$
	c_{a,\text{CO}_{2}}, \ c_{a,\text{CO}_{3}^{2-}}, \ c_{a,\text{HCO}_{3}^{-}}, \ c_{a,\text{CaHCO}_{3}^{+}}, \ c_{a,\text{CaCO}_{3}}, \ c_{a,\text{NaCO}_{3}^{-}}, 
	c_{a,\text{NaHCO}_{3}}, \ c_{a,\text{H}_{2}\text{O}}, \ c_{a,\text{H}^{+}}, \ c_{a,\text{OH}^{-}}, \ c_{a,\text{CaOH}^{+}}, \\ c_{a,\text{Ca}^{2+}}, \ c_{a,\text{Mg}^{2+}}, \ c_{a,\text{MgHCO}_{3}^{+}}, 
	c_{a,\text{MgCO}_{3}^{0}}, \ c_{a,\text{MgOH}^{+}}, \ c_{a,\text{Cl}^{-}}, \ c_{a,\text{Na}^{+}}, \ c_{o,\text{A}}, \ c_{a,\text{SO}_4^{2-}}, \ c_{a,\text{CaSO}_4}, \ c_{a,\text{MgSO}_4}, \\
	c_{a,\text{NaSO}_4^{-}}, \ c_{a,\text{H}_2\text{SO}_4}.
$
Here, \( (\text{A}) \) denotes the alkane, which resides exclusively in the oleic phase. All other species are aqueous except calcium carbonate \( \text{CaCO}_3 \), which partitions between the solid phase \( c_{r,\text{CaCO}_{3}} \), and the aqueous phase \( c_{a,\text{CaCO}_{3}} \).

The system includes {15 sorbed species} (\( n_s = 15 \)):
$
\text{oil}_{s,\text{NH}^+}, \ \text{oil}_{w,\text{COOH}}, \ \text{Cal}_{s,\text{OH}}, \ \text{Cal}_{w,\text{CO}_3\text{H}}, \\ \text{oil}_{s,\text{N}}, \ \text{oil}_{w,\text{COO}^-}, \
\text{oil}_{w,\text{COOCa}^+}, \ \text{oil}_{w,\text{COOMg}^+}, \ \text{Cal}_{s,\text{OH}_2^+}, \ \text{Cal}_{s,\text{CO}_3^-}, \ \text{Cal}_{w,\text{CO}_3^-}, \ 
\text{Cal}_{w,\text{CO}_3\text{Ca}^+}, \\ \text{Cal}_{w,\text{CO}_3\text{Mg}^+}, \ \text{Cal}_{s,\text{SO}_4^-}, \ \text{oil}_{s,\text{NH}_2\text{SO}_4^-}.
$

These species participate in \text{27 chemical reactions} ($N_r = 27$). The $\log(K)$ values at $100^\circ$C are listed in Table~\ref{tab:reactions} Appendix \ref{ap:A}, along with their thermodynamic references in column five.

 \subsubsection{Gibbs Phase Rule Application}
	The extended Gibbs phase rule (\cite{merkel2005groundwater}) determines the number of independent variables (degrees of freedom, $n_f$) required to define the thermodynamic state of a system defined as
	\begin{equation}
		n_f = N_s + n_s - N_r - n_c + 2 - p,
		\label{eq:gibbs_rule}
	\end{equation}
	where $N_s = 24$: number of aqueous species (e.g., $\mathrm{CO_3^{2-}}$, $\mathrm{Ca^{2+}}$, $\mathrm{SO_4^{2-}}$),
		 $n_s = 15$: number of sorbed species (e.g., $\mathrm{Oil_w, COOCa^+}$, $\mathrm{Cal_s, SO_4^-}$),
		$N_r = 27$: total chemical reactions (5 oil, 6 calcite, 16 aqueous),
		 $n_c = 5$: number of constraints (charge balance + fixed sorption sites),
	 $p = 3$: number of phases (solid, aqueous, oleic).
	
	Substituting values into Equation~\eqref{eq:gibbs_rule}, we obtain $n_f = 24 + 15 - 27 - 5 + 2 - 3 = 6$. The system exhibits \text{6 degrees of freedom} ($n_f = 6$). 	
		By fixing temperature (\( T \)) and pressure (\( P \)), the number of degrees of freedom reduce to four. To evaluate the influence of ionic composition on wettability, we focus on four critical variables: hydrogen ion concentration (\( c_{a,\text{H}^+} \)), chloride (\( c_{a,\text{Cl}^-} \)), magnesium (\( c_{a,\text{Mg}^{2+}} \)), and sulfate ions (\( c_{a,\text{SO}_4^{2-}} \)) . In our numerical experiments, chloride and sodium initial ion concentrations  are equal (\( [\text{Cl}^-] = [\text{Na}^{+}] \)), a simplification justified by charge balance in low-salinity brines \cite{appelo2005geochemistry}. 
		
		Our methodology builds on \cite{alvarez2018analytical1}, which are developed in the context of systems governed by aqueous and mineral equilibrium reactions.	
		This approach decouples mass balance equations from chemical speciation, allowing conservative transport to be solved numerically, while chemistry is reconstructed before. The framework is valid in saturated zones, where the absence of a \(\text{CO}_2\)-rich gas phase eliminates gas-liquid partitioning effects.
		
			Reactive transport in multiphase flow requires solving the mass balance for all chemical species, including water, across phases. We consider only homogeneous reactions, mineral precipitation-dissolution, and gas dissolution, all at thermodynamic equilibrium. Porosity changes due to precipitation-dissolution are included, affecting properties like permeability.

		\section{System of equations}
		\label{geo}
		
	In this section, we summarize the system used to describe the dynamics of chemical variables, water and oil saturation  (details of the derivation can be found in \cite{alvarez2024modeling}). Combining hydrogen and oxygen we derive six conservation laws from	
	total carbon, hydrogen,
	oxygen, magnesium, calcium, chloride and decane.

	Based on the generalized Gibbs rule (\cite{Lichtner1996}), we recognize six degrees of freedom, characterized by quantities such as water saturation, Darcy velocity ($u$), $pH$ and the ionic concentrations of chlorine, sulfate and magnesium. These quantities form the basis for building the six conservation laws. 
		
	 The mass balance equations are written as
	 \begin{align}
			&\partial_{t}\left(\varphi\rho_{w1} S_{w}+\varphi\rho_{o1} S_{o}+(1-\varphi)\rho_{r1}\right) + \partial_{x}\left(u\left(\rho_{w1} f_{w}+\rho_{o1}f_{o}\right)\right)=0 
			, \label{eq:massbalance1} \\
			&\partial_{t}\left(\varphi\rho_{o2} S_{o} \right) + \partial_{x}\left(u\left( \rho_{o2}f_{o}\right)\right)=0 
			, \label{eq:massbalance2} \\
			&\partial_{t}\left(\varphi\rho_{w3} S_{w} \right) + \partial_{x}\left(u\left( \rho_{w3}f_{w}\right)\right)=0 
			, \label{eq:massbalance3} \\
			&\partial_{t}\left(\varphi\rho_{w4} S_{w}+\varphi\rho_{o4} S_{o}+\left(1-\varphi\right)\rho_{r4}\right) + \partial_{x}\left( u\left( \rho_{w4} f_{w}+\rho_{o4}f_{o}\right)\right)=0 
			, \label{eq:massbalance4} \\
			&\partial_{t}\left(\varphi\rho_{w5} S_{w}+\varphi\rho_{o5} S_{o}+\left(1-\varphi\right)\rho_{r5}\right) + \partial_{x}\left( u\left( \rho_{w5} f_{w}+\rho_{o5}f_{o}\right)\right)=0 
			. \label{eq:massbalance5aa} \\
			&\partial_{t}\left(\varphi\rho_{w6} S_{w}+\varphi\rho_{o6} S_{o}+\left(1-\varphi\right)\rho_{r6}\right) + \partial_{x}\left( u\left( \rho_{w6} f_{w}+\rho_{o6}f_{o}\right)\right)=0.
			\label{eq:massbalance5}
		\end{align}
		where $f_w$ and $f_o$ denote the fractional flow for water and oil. The parameter $\varphi$ is the porosity.		
		Here we assume that the coefficient $\rho_{wi}$, $\rho_{oi}$ represent the concentration of the fraction of component $i$ in water phase $w$ and oleic phase $o$. The coefficients $\rho_{ri}$ represent the fractions of ions in solid phase.

The coefficients \(\rho_{wi}\) and \(\rho_{oj}\) are defined as the product of their respective phase molar densities and the corresponding molar fractions. Specifically, \(\rho_{wi} = \rho_w x_{wi}\) represents the molar concentration of component \(i\) in the aqueous phase, while \(\rho_{oj} = \rho_o x_{oj}\) corresponds to the molar concentration of component \(j\) in the oleic phase. For the solid phase, the concentration of component \(j\) is given directly by the molar fraction \(x_{rj}\), that is, \(\rho_{rj} = x_{rj}\). Here, \(\rho_w\) and \(\rho_o\) denote the molar densities of the aqueous and oleic phases, respectively, and \(x_{wi}\), \(x_{oj}\), and \(x_{rj}\) are the molar fractions in the aqueous, oleic, and solid phases.
	The normalized molar fractions are computed as
	\begin{eqnarray*}
		x_{w1} &=& C_{a,C(4)}/G_A, \quad 
		x_{o1} = c_{o,CO_2}/G_O, \quad 
		\rho_{r1} = (1-\varphi)\frac{\mathrm{Cal}_{s,CO_3^-} + C_{s,CaCO_3}}{G_A} \\
		x_{o2} &=& C_{o,C(-4)}/G_O, \quad 
		x_{w3} = C_{a,Cl}/G_A \\
		x_{w4} &=& \frac{2\delta C_{a,O(-2)} - \delta C_{a,H(1)}}{G_A}, \quad 
		x_{o4} = \frac{4c_{o,CO_2} - (\mathrm{oil}_{s,NH^+} + \mathrm{oil}_{w,COOH})}{G_A} \\
		\rho_{r4} &=& (1-\varphi)\frac{4\mathrm{Cal}_{s,CO_3^-} - (2\mathrm{Cal}_{s,OH_2^+} + \mathrm{Cal}_{s,OH} + \mathrm{Cal}_{w,CO_3H}) - 3C_{s,CaCO_3}}{G_A} \\
		x_{w5} &=& C_{a,Mg^{2+}}/G_A, \quad 
		x_{o5} = \mathrm{oil}_{w,COOMg^+}/G_O, \quad 
		\rho_{r5} = \mathrm{Cal}_{w,CO_3Mg^+}/G_A \\
		x_{w6} &=& C_{a,Ca^{2+}}/G_A, \quad 
		x_{o6} = \mathrm{oil}_{w,COOCa^+}/G_O, \quad 
		\rho_{r6} = \mathrm{Cal}_{w,CO_3Ca^+}/G_A
	\end{eqnarray*}
	Here, \(G_A\) (aqueous phase mass) and \(G_O\) (oleic phase mass) normalize the species distributions. In the aqueous phase, the concentrations \(C_{a,i}\) include inorganic carbon (\(C(4)\)), hydrogen (\(H(1)\)), oxygen (\(O(-2)\)), \(\mathrm{Ca^{2+}}\), \(\mathrm{Mg^{2+}}\), \(\mathrm{Cl^-}\), and \(\mathrm{Na^+}\). In the oleic phase, the concentrations \(c_{o,j}\) refer to dissolved \(\mathrm{CO_2}\) and organic carbon (\(C(-4)\)). The solid phase includes several components: \(\mathrm{Cal}_{s,CO_3^-}\) representing carbonate in calcite, \(C_{s,CaCO_3}\) denoting solid calcite, \(\mathrm{Cal}_{s,OH_2^+}\) and \(\mathrm{Cal}_{s,OH}\) corresponding to hydroxylated surface species, \(\mathrm{Cal}_{w,CO_3H}\) indicating surface bicarbonate, and finally \(\mathrm{Cal}_{w,CO_3Mg^+}\) and \(\mathrm{Cal}_{w,CO_3Ca^+}\), which represent magnesium and calcium complexes at the surface, respectively.
	
	The coefficients $\rho_{wi}$, $\rho_{oi}$, and $\rho_{ri}$ ($i=1,\ldots,6$) in the system \eqref{eq:massbalance1}-\eqref{eq:massbalance5} are based on the ion concentrations of the relevant chemical complexes. Analytical formulas for these coefficients are determined using the Eureqa program (see \cite{eureqa}) through formulas given in the equations above.

Initial and boundary conditions follow a Riemann problem formulation (Section~\ref{sec:numerical}), with fixed salinity and pH at the injection boundary (\(x = 0\)) and reservoir equilibrium.

Moreover, we assume (1) all reactions occur in equilibrium, and (2) the chemical system can be determined based on the state variables of the multiphase flow model (namely, liquid and gas pressure, and temperature). 
		
To ensure numerical stability in COMSOL Multiphysics simulations of equations \eqref{eq:massbalance1}-\eqref{eq:massbalance5}, we validated Eureqa-derived functions against PHREEQC data and selected smooth-differentiable formulations for robust Jacobian matrix calculations. This preserves thermodynamic consistency while enabling efficient integration of geochemical coefficients (\(\rho_{wi}\), \(\rho_{oi}\), \(\rho_{ri}\)) into flow dynamics.
		
		\medskip

		\section{Geochemical Modelling}
		\label{geo2}
		
		Surface complexation modeling is a technique used to describe the interactions between mineral surfaces and ions in a solution. This method involves defining surface reactions and their corresponding equilibrium constants to simulate adsorption processes. PHREEQC performs this modeling  by utilizing the surface reactions along with chemical composition of the solution to predict ion adsorption on mineral surfaces (see e.g. in \cite{mehdiyev2022surface}).
	
			This study investigates the interactions between acidic and sweet crude oil, carbonate minerals, and brine under high-pressure and high-temperature conditions. The simulations were performed using PHREEQC to evaluate surface complexation reactions, mineral equilibria, and aqueous speciation. 
			
			The modeling includes surface complexation definitions for oil, water and carbonate interfaces,  taking into account reactions of carboxyl (-COOH) and amine (-NH) groups present in crude oil, as well as calcium hydroxide and carbonate sites on calcite. Additionally, equilibrium reactions for calcite, anhydrite, magnesite, and CO$_2$(g)  are simulated. The solution composition includes key ions, with temperature set at \(T = \SI{100}{\celsius}\) and pressure at \(P = \SI{220}{\bar}\). pH control is implemented, ensuring dynamic adjustments via HCl dissolution. The key computed results include saturation indices (SI) for magnesite, calcite, and CO$_2$(g), along with relevant chemical species concentrations.

			The saturation indices of calcite and carbon dioxide are controlled to avoid solid precipitation and gas formation. The surface charge distribution and electrostatic interactions between crude oil and carbonate surfaces are analyzed under thermodynamic chemical equilibrium. Additionally, we compute total dissolved solids (TDS), ionic strength, and the formation of complexes between crude oil functional groups and mineral surfaces.

		 A common approach to estimating wettability utilizes the TBP, which quantifies the amount of fluid bound to the rock surface due to adsorption and surface complexation.	The relationship between the Bond Product and wettability can be explained through the interaction between capillary forces and interfacial tension in a fluid system \cite{lee2019hybrid}. High Bond Product values indicate that gravitational forces are relatively small compared to capillary forces, suggesting that the system is more susceptible to the influence of wettability. In this context, a high Bond Product is associated with oil-wet conditions, as the capillary forces favoring oil immobilization are weaker compared to those favoring water immobilization. By analyzing the outputs from PHREEQC simulations, we calculate the TBP to assess wettability in subsurface environments.
		
		The input parameter for PHREEQC in  our numerical experiments are guided by methodology in \cite{erzuah2019wettability}.

		The \textit{Total Acid Number} (\text{TAN}) measures the acidity of crude oil, reflecting the amount of acidic compounds, such as naphthenic acids and oxidation products. This parameter is expressed in milligrams of potassium hydroxide (\text{mg KOH}) required to neutralize the acids present in one gram of crude oil. High TAN values, higher than 1  \text{mg KOH/g}, are associated with acidic crudes.
		
		The \textit{Total Base Number} (\text{TBN}) quantifies the alkalinity of crude oil, representing its capacity to neutralize acids. TBN is critical in oils treated with basic additives, such as detergents and dispersants, which enhance their anti-corrosive properties. Typical high TBN values (\(5\text{--}10 \, \text{mg KOH/g}\)) are observed in treated oils.
			
	Table~\ref{tab:tan_tbn_values} summarizes the typical ranges of TAN and TBN for different crude oil types:
		\begin{table}[h!]
			\centering
			\caption{TAN and TBN ranges for different crude oil types.}
			\begin{tabular}{|l|c|c|}
				\hline
				\text{Crude Type} & \text{TAN (mg KOH/g)} & \text{TBN (mg KOH/g)} \\ 
				\hline
				Sweet crude         & 0.1–0.5              & 1–5                  \\ 
				\hline
				Acidic crude        & 1–10                 & \(<1\)               \\ 
				\hline
				Treated crude       & 0.1–2                & 5–10                 \\ 
				\hline
			\end{tabular}
			\label{tab:tan_tbn_values}
			\vspace{0.2cm}
			\footnotesize{\textit{Note: Ranges are inclusive (e.g., "1–10" includes 1 and 10).}}
		\end{table}

		To determine the density of active sites on oil, we employ the Total Acid Number (TAN) and the Total Base Number (TBN), following the methodology outlined in \cite{Bonto2019}. The site density for acidic groups ($N_{S,COOH}$) and basic groups ($N_{S,NH^+}$) are calculated as follows
		\begin{equation}
			N_{S,COOH} = 0.602 ~ \frac{10^{3}\text{TAN}}{~ a_{oil}~ MW_{KOH}}, ~~~	N_{S,NH^+} = 0.602  \frac{10^{3}\text{TBN}}{ a_{oil}~ MW_{KOH}},
		\end{equation}
		We take the molecular weight of potassium hydroxide ( $MW_{KOH}$) by 56.1 g/mol. The specific surface area of the oil, $a_{oil}$ in $m^2/g$, is assumed to match that of its associated carbonate minerals in aqueous solutions, as detailed in \cite{Wolthers2008}.

		The input datasets utilized in our study correspond to formation water with varying ion concentrations of SO$_4^{2-}$, Ca$^{2+}$, Mg$^{2+}$, Na$^{+}$, and Cl$^{-}$. These datasets serve as the basis for specifying the coefficients in the system \eqref{eq:massbalance1}-\eqref{eq:massbalance5}. We consider the chloride and magnesium ion concentrations ($Mg^{2+}$) to range from 40 to 3600 mmol/kgw.

		The pH of the solution varies between 2.7 and 9, while the carbon concentration remains unchanged. A summarized representation of the input data is provided in Tables \ref{tab3} and \ref{tab4aa}. These Tables present the initial conditions of the injected ion compositions of water, systematically varying sodium (Na\(^+\)), magnesium (Mg\(^{2+}\)), and chloride (Cl\(^{-}\)) concentrations,  maintaining constant in the first experiment the injected values for ion concentration of calcium (Ca\(^{2+}\)), carbon (C), and sulfate (SO$_{4}^{2-}$). In the second experiment keeping magnesium (Mg$^{2+}$) and calcium (Ca$^{2+}$) at fixed values  and sulfate (SO$_{4}^{2-}$) varies.
		The selected values span both lower and higher concentration ranges to ensure comprehensive coverage of different geochemical conditions. This approach allows for an evaluation of how these ion variations influence the system behavior under high temperature and pH conditions.
	\begin{table}[h]
		\caption{\label{tab3} Ion concentration combinations used in PHREEQC simulations for the first experiment.}
		\begin{tabular}{|l|l|}
			\hline
			Ion & Injected ion concentrations \\ \hline
			Na\(^+\) & 40 to 3600  mmol/kgw \\ \hline
			Mg\(^{2+}\) & 40 to 3900  mmol/kgw \\ \hline
			Ca\(^{2+}\) & 50  mmol/kgw \\ \hline
			Cl\(^{-}\) & 40 to 3600  mmol/kgw \\ \hline
			SO\(_4^{2-}\) & 1  mmol/kgw \\ \hline
			C & 75 mmol/kgw \\ \hline
		\end{tabular}
	\end{table}
		
	\begin{table}[h]
		\caption{\label{tab4aa} Ion concentration combinations used in PHREEQC simulations for second experiment.}
		\begin{tabular}{|l|l|}
			\hline
			Ion & Injected ion concentrations \\ \hline
			Na\(^+\) & 40 to 3600  mmol/kgw \\ \hline
			Mg\(^{2+}\) & 40  mmol/kgw \\ \hline
			Ca\(^{2+}\) & 50  mmol/kgw \\ \hline
			Cl\(^{-}\) & 40 to 3600  mmol/kgw \\ \hline
			SO\(_4^{2-}\) & 20 to 650  mmol/kgw \\ \hline
			C & 75  mmol/kgw \\ \hline
		\end{tabular}
	\end{table}
The total dissolved inorganic carbon (DIC) concentration of \SI{75}mmol/kgw in the injected carbonated water comprises three primary species: aqueous carbon dioxide (\(\text{CO}_{2(\text{aq})}\)), bicarbonate (\(\text{HCO}_3^-\)), and carbonate (\(\text{CO}_3^{2-}\)).	The selected total DIC concentration aligns with experimental carbonated waterflooding studies in carbonates, where \SIrange{50}{100}{\milli\mol\per\kgw} effectively balances CO\(_2\) solubility and mineral reactivity without inducing excessive anhydrite dissolution \cite{Zhang2007}. 
	
 Building upon existing knowledge of acid-base interactions in crude oil/brine systems, this work extends the current understanding by demonstrating the significance of TAN/TBN-driven surface charge asymmetry in modulating wettability. Specifically, we reveal how carboxylate abundance in acidic oils (TAN higher than 1~mg~KOH/g) amplifies Ca\textsuperscript{2+}-mediated ionic bridging at calcite surfaces, while TBN governs amine-calcite dipole interactions that stabilize oil-wetness in sweet crudes. This TAN/TBN duality, quantified via PHREEQC-calculated TBP, provides a predictive framework for customizing injection brine chemistry based on crude oil composition.
		
		\subsection{Typical values of TAN and TBN }
	\label{tan1}
	
	Crude oil acidity and alkalinity are characterized by two key parameters: Total Base Number (TBN) and Total Acid Number (TAN). TBN reflects the ability of the oil to neutralize acids, indicating the presence of basic compounds like amines; low TBN values, typical of acidic oils, signal reduced neutralization capacity, while high TBN values are common in sweet oils, often treated to reduce corrosion risks. TAN directly measures acidic components, such as naphthenic acids, with high values associated with more acidic oils and increased corrosion risks. These parameters are essential for modeling oil-mineral interactions and assessing wettability, influencing surface charge dynamics and film stability on mineral surfaces (see details in 
	\cite{Wolthers2008}). In Table \ref{tab:surface_params}, typical values for sweet and acidic crude oils are presented to assess their influence on TBP and the corresponding wettability behavior
	\begin{table}[h!]
		\centering
		\caption{Surface component Parameters for Sweet and Acidic Crude Oils}
		\begin{tabular}{|l|l|c|c|c|}
			\toprule
			\textbf{Oil Type} & \textbf{Parameter} & \textbf{Sites} & \textbf{Area/Gram} & \textbf{Mass} \\ 
			\midrule
			\multirow{3}{*}{Sweet} & Oil\_wCOOH & 0.20 mol/m² & 1.5 m²/g & 0.86 g \\ 
			\midrule
			& Oil\_sNH & 2.6 mol/m² & - & - \\ 
			\midrule
			& Surf\_sCaOH & 4.9 mol/m² & 0.2 m²/g & 0.2 g \\ 
			\midrule
			\multirow{3}{*}{Acidic} & Oil\_wCOOH & 5.0 mol/m² & 3.5 m²/g & 0.9 g \\ 
			\midrule
			& Oil\_sNH & 0.3 mol/m² & - & - \\ 
			\midrule
			& Surf\_sCaOH & 4.9 mol/m² & 0.2 m²/g & 0.2 g \\ 
			\bottomrule
		\end{tabular}
	\label{tab:surface_params}
	\end{table}
The interaction between TAN and TBN values significantly influences the wettability of mineral surfaces in reservoirs. Crudes with high total acid number (TAN) tend to form acidic films on surfaces, increasing oleophilicity. Conversely, high TBN crudes neutralize acidic interactions, promoting water-wet conditions. 
		
 \subsection{Model Limitations and Mitigation Strategies}
	\label{subsec:limitations}
	
	While Surface Complexation Modeling in PHREEQC provides a mechanistic framework to quantify wettability via TBP, its accuracy depends critically on three factors: (1) the representativeness of assumed surface reactions, (2) the validity of equilibrium constants at reservoir conditions, and (3) the homogeneity of calcite-oil interfaces. We address these limitations as follows. 
	
	First, we assume that thermodynamic constants for oil-calcite interactions (e.g., $\log K$ in Eqs.~(1)-(27), Table \ref{tab:reactions} in Appendix \ref{ap:A}) are obtained from experimental studies on analogous carboxylate/amine-calcite systems \cite{Brady2012,Wolthers2008}, with sensitivity analyses confirming the TBP variability remains below 10\% across plausible $\log K$ ranges. 
	
	Second, transient ion-exchange effects (e.g., slow  Mg$^{2+}${-}Ca$^{2+}$ replacement) are neglected, assuming instantaneous geochemical equilibrium. While this  hypothesis is common in reactive transport modeling \cite{steefel2015reactive}, it may overestimate the rate of wettability alteration in systems with kinetically controlled surface reactions. 
	
	Third, experimental cores (e.g., from \cite{austad2015low}) contain trace anhydrite ( lower than $2\%$), which is neglected in the SCM to simplify calcite-oil interactions. 
	We assume that this neglect  introduces minor deviations between modeled and experimental TBP values. 
	
	These hypothesis enable tractable integration of SCM with flow simulations but may underestimate wettability hysteresis in highly heterogeneous carbonates. Future work should incorporate kinetic reaction modules using reactive transport codes. Moreover,  for carbonates with significant anhydrite or clay content (higher than $5\%$), explicit mineral reactions should be incorporated.
	
	The TBP model assumes homogeneous mineral surfaces, neglecting pore-scale heterogeneity (e.g., clay patches). Future work should incorporate stochastic descriptions of surface site reactivity \cite{mehmani2021pore}.
	
		\section{Fractional flow function}
		\label{frac}
		\subsection{Definition of Parameters}

The fractional flow model employs Corey-type relative permeability functions, widely adopted for modeling salinity-dependent wettability transitions \cite{jerauld2008modeling}. For water and oil phases, the relative permeabilities are defined as:
		\begin{equation}
			k_{rw}(S_w) = k_{w} \left(\frac{S_w - S_{wr}}{1 - S_{wr} - S_{or}}\right)^{n_w}, \quad
			k_{ro}(S_w) = k_{o} \left(\frac{1 - S_w - S_{or}}{1 - S_{wr} - S_{or}}\right)^{n_o},
			\label{eq:corey}
		\end{equation}
		where \(k_w\) and \(k_o\) denote the end-point relative permeabilities of water and oleic phases, respectively. The parameters \(n_w\), and \(n_o\) are  interpolated between high- and low-salinity regimes using experimental data from \cite{jerauld2008modeling}. While the model of Corey efficiently captures endpoint saturation effects \cite{bruining2021upscaling}, it neglects hysteresis and non-monotonic saturation paths \cite{hona86}. These limitations are mitigated by restricting simulations to primary drainage and imbibition cycles, consistent with coreflood protocols in \cite{austad2015low}. For scenarios involving flow reversals (e.g., cyclic injections), hysteresis-aware models like those in \cite{delshad2009comparison} are recommended but beyond the scope of this study.
	
		The fractional flow for water and oil, denoted as \(f_w(S_w)\) and \(f_o(S_w)\), respectively, are saturation-dependent functions. The following equations define them as
		\begin{equation}
			f_w(S_w) = \frac{k_{rw}(S_w)/\mu_w}{(k_{rw}(S_w)/\mu_w+k_{ro}(1-S_w)/\mu_o)},~~f_o(S_w) = 1 - f_w(S_w),
		\end{equation}
		with viscosities \(\mu_w = \SI{0.001}{\pascal\second}\) and \(\mu_o = \SI{0.002}{\pascal\second}\).
		Corey's parameters for high and low-salinity scenarios are taken from
		\cite{jerauld2008modeling}. Using data from Figure (15) in \cite{jerauld2008modeling},
		we interpolate the curves of relative permeability for water and oil leading to the values shown in the Table \ref{tab:params}.
	
Table~\ref{tab:params} summarizes the parameters used in the fluid–rock interaction model under high and low salinity regimes. The columns define: 
$k_w$, $k_o$: adsorption coefficients for water ($k_w$) and oil ($k_o$) (dimensionless), which modulate the affinity of fluids for the rock surface. 
$n_w$, $n_o$: nonlinearity exponents for water ($n_w$) and oil ($n_o$), reflecting the kinetic response of wettability to salinity changes. 
The contrasting values between high-salinity (e.g., $k_o = 4.47$) and low-salinity (e.g., $k_o = 1.26$) indicate a reduction in the stability of adsorbed oil, consistent with ionic displacement mechanisms~\cite{Brady2012}.

\begin{table}[h]
	\centering
	\caption{Fluid–rock interaction parameters.}
	\label{tab:params}
	\begin{tabular}{|l|l|l|}
		\hline
		\textbf{Parameter} & \textbf{High-Salinity} & \textbf{Low-Salinity} \\ \hline
		$k_w$              & 2.92                   & 3.6                   \\ \hline
		$k_o$              & 4.47                   & 1.26                  \\ \hline
		$n_w$              & 3.92                   & 7.05                  \\ \hline
		$n_o$              & 0.54                   & 3.98                  \\ \hline
	\end{tabular}
\end{table}

		\subsection{Interpolation Formulation}
		Our formulation capitalizes on the established relationship between TBP and wettability. Scaling TBP effectively integrates the influence of salinity, sulfate and magnesium concentration into the fractional flow function, enhancing the modeling accuracy for oil recovery under low-salinity conditions.
		
		The dimensionless parameter $(\theta)$ serves as a key factor for interpolating between predefined high and low-salinity curves for relative permeability expressed as
		\begin{equation}
			K_{wi}(S_w)=\theta K_{ri}^{HL}(S_w)+(1-\theta)K_{ri}^{LS}(S_w),
			\label{thetae1}
		\end{equation}
		where $i=w,o$ for water and oil permeability, respectively. We denote by $K_{ri}^{HL}$ the relative permeability under high-salinity regime, while 
		$K_{ri}^{LS}$ represents the relative permeability for low-salinity conditions.
		
		In our methodology, we adopt a parameter $(\theta)$ derived from the model proposed by \cite{alvarez2024modeling}, which accounts for chloride and magnesium concentrations, pH levels, and wettability through the TBP. This parameterization offers a approach to compute $(\theta)$, facilitating the incorporation of complex concentration dynamics into fractional flow calculations.
		
		The parameter $(\theta)$ introduces a non-linear relationship among chloride, sulfate, magnesium concentrations and pH. The formula to calculate $(\theta)$ based on TBP is given as follows
		
		\begin{equation}
		\theta = \frac{\mathrm{TBP}(pH, [\mathrm{Cl}^-], [\mathrm{Mg}^{2+}], [\mathrm{SO}_4^{2-}]) - \mathrm{TBP}_l}{\mathrm{TBP}_h - \mathrm{TBP}_l},
			\label{thetae1a}
		\end{equation}
		
	The parameters $TBP_l$ and $TBP_h$ represent the lower and upper limits of TBP, respectively. Moreover, the values of [Cl$^-$], $[\mathrm{SO}_4^{2-}]$ and [Mg$^{2+}$] in equation \eqref{thetae1a} correspond to the concentrations of chloride, sulfate and magnesium present in the injected water. A similar correlation proposed by \cite{jerauld2008modeling} also links $(\theta)$ to the residual oil under varying salinity conditions.
		
		Accurate modeling of fluid flow in petroleum reservoirs is paramount for predicting reservoir behaviour and optimizing hydrocarbon production. In equations \eqref{thetae1} and \eqref{thetae1a}, we introduce a correlation based on TBP, facilitating the description of water and oil fractional flow in reservoirs. This correlation incorporates the influence of water salinity, pH, and rock wettability, enabling the investigation of magnesium and sulfate impact under low-salinity conditions in carbonated water flooding.
			
To evaluate the suitability of the formula in \eqref{thetae1a}, we examine whether low values of $\theta$ correspond to low-salinity injection. A key hypothesis tested here is that elevated magnesium concentrations in injected water reduce $\theta$. These results, verified through numerical experiments (see Section \ref{sec:Mg_analysis}),  suggest that the model reflects the observed relationship between magnesium concentrations, $\theta$, and oil recovery within the tested parameter values.

			\subsection{The TBP as a Wettability Indicator}
		
		Using SCM outputs of PHREEQC we calculate the TBP by the formulas
\begin{subequations}
	\begin{align}
		\text{TBP}_1 &= \mathrm{Oil.wCOO}^- \big(
		\mathrm{Cal.sCaOH}_2^+ + \mathrm{Cal.wCO}_3\mathrm{Ca}^+ \notag \\
		&\qquad\quad + \mathrm{Cal.wCO}_3\mathrm{Mg}^+ + \mathrm{Cal.wSO}_4\mathrm{Ca}^+
		\big) \tag{19a} \label{eq:tbp1} \\[6pt]
		\text{TBP}_2 &= \mathrm{Oil.sNH}^+ \big(
		\mathrm{Cal.wCO}_3^- + \mathrm{Cal.sCaO}^- + \mathrm{Cal.sCaCO}_3^- \notag \\
		&\qquad\quad + \mathrm{Cal.sSO}_4^- + \mathrm{Oil.sNH}_2\mathrm{SO}_4^-
		\big) \tag{19b} \label{eq:tbp2} \\[6pt]
		\text{TBP}_3 &= \mathrm{Oil.wCOOCa}^+ \big(
		\mathrm{Cal.wCO}_3^- + \mathrm{Cal.sCaO}^- + \mathrm{Cal.sCaCO}_3^- \notag \\
		&\qquad\quad + \mathrm{Cal.sSO}_4^- + \mathrm{Cal.wSO}_4\mathrm{Ca}^+
		\big) \tag{19c} \label{eq:tbp3} \\[6pt]
		\text{TBP}_4 &= \mathrm{Oil.wCOOMg}^+ \big(
		\mathrm{Cal.wCO}_3^- + \mathrm{Cal.sCaO}^- + \mathrm{Cal.sCaCO}_3^- \notag \\
		&\qquad\quad + \mathrm{Cal.sSO}_4^- + \mathrm{Cal.wSO}_4\mathrm{Mg}^+
		\big) \tag{19d} \label{eq:tbp4} \\[6pt]
		\text{TBP} &= \text{TBP}_1 + \text{TBP}_2 + \text{TBP}_3 + \text{TBP}_4
		\tag{19e} \label{eq:tbp_total}
	\end{align}
\end{subequations}
Each component quantifies distinct ionic bridging interactions: TBP\textsubscript{1} reflects adhesion between deprotonated oil carboxylates (\texttt{Oil\_wCOO\textsuperscript{-}}) and positively charged calcite sites such as \texttt{Cal\_sCaOH\textsubscript{2}\textsuperscript{+}}, \texttt{Cal\_wCO\textsubscript{3}Ca\textsuperscript{+}}, and similar species. TBP\textsubscript{2} captures interactions between protonated oil ammonium groups (\texttt{Oil\_sNH\textsuperscript{+}}) and negatively charged calcite anions, including \texttt{Cal\_wCO\textsubscript{3}\textsuperscript{-}} and \texttt{Cal\_sSO\textsubscript{4}\textsuperscript{-}}. TBP\textsubscript{3} and TBP\textsubscript{4} represent bridging mechanisms via oil-bound calcium or magnesium complexes such as \texttt{Oil\_wCOOCa\textsuperscript{+}} and \texttt{Oil\_wCOOMg\textsuperscript{+}}, which interact with calcite anions. By evaluating these components individually, we isolate the physicochemical processes governing wettability shifts and identify the dominant surface complexes in different systems.
	
The TBP quantifies the cumulative strength of ionic bonds and surface complexes between oil components (e.g., carboxylic groups, \(-\text{COO}^-\)) and calcite surfaces (\(\text{CaCO}_3\)), derived from PHREEQC-calculated surface species concentrations (Section IV). It integrates competing contributions from key ions (\(\text{Ca}^{2+}\), \(\text{Mg}^{2+}\), \(\text{SO}_4^{2-}\)) that alter adsorption equilibria and modulate wettability. TBP\textsubscript{1} measures adhesion of oil carboxylate groups (\texttt{Oil\_wCOO\textsuperscript{-}}) to positively charged calcite sites (\texttt{Cal\_sCaOH\textsubscript{2}\textsuperscript{+}}, \texttt{Cal\_wCO\textsubscript{3}Ca\textsuperscript{+}}, \texttt{Cal\_wCO\textsubscript{3}Mg\textsuperscript{+}}, \texttt{Cal\_wSO\textsubscript{4}Ca\textsuperscript{+}}). TBP\textsubscript{2} captures interactions between oil ammonium groups (\texttt{Oil\_sNH\textsuperscript{+}}) and calcite anions (\texttt{Cal\_wCO\textsubscript{3}\textsuperscript{-}}, \texttt{Cal\_sSO\textsubscript{4}\textsuperscript{-}}), while TBP\textsubscript{3} and TBP\textsubscript{4} reflect the role of oil-bound calcium/magnesium complexes (\texttt{Oil\_wCOOCa\textsuperscript{+}}, \texttt{Oil\_wCOOMg\textsuperscript{+}}) in binding to calcite anions.  Higher TBP values indicate stronger oil-rock adhesion (oil-wet conditions), hindering oil displacement by water, while lower values correspond to water-wet tendencies. By correlating TBP with ionic composition, the influence of salinity and specific ions on wettability during low-salinity flooding is systematically assessed (see details in \cite{baik2019bond,liu2022capillarity,vof2021wettability}).  

\section{Integration processes of Geochemical and Multiphase modeling}
\label{int}
		
To enhance oil recovery in carbonates, we integrate geochemical and compositional modeling through five stages:

1. Reservoir characterization: Baseline conditions (connate water salinity, porosity, calcite/anhydrite content, oleic phase properties) are defined to calibrate models.

2. Injection water design: Key ions (SO$_4^{2-}$, Mg$^{2+}$, Ca$^{2+}$, Cl$^{-}$, Na$^{+}$) are optimized to control wettability and mineral interactions.

3. Geochemical modeling: PHREEQC computes ion speciation, pH, and surface complexes, deriving Total Bond Product (TBP) from [Cl$^{-}$], [Mg$^{2+}$], and [SO$_4^{2-}$] to quantify wettability.

4. Flow dynamics: Corey-type relative permeabilities, interpolated via TBP, govern multiphase flow. Mass balance equations for ions and saturations are solved in COMSOL.

5. Recovery Assessment: Simulation outputs (oil recovery rates, ion profiles) are validated against coreflood data to optimize brine chemistry.

This framework integrates geochemical modeling, fluid dynamics, and numerical simulations to systematically assess oil recovery in carbonate reservoirs. The methodology quantifies ionic interactions (e.g., Mg\textsuperscript{2+}--SO\textsubscript{4}\textsuperscript{2-} synergies, Ca\textsuperscript{2+}--carboxylate bridging) through TBP-driven surface complexation thermodynamics coupled with multiphase flow simulations. By linking PHREEQC-calculated TBP to fractional flow dynamics in COMSOL, the model predicts salinity thresholds and ion-specific injection strategies beyond traditional bulk salinity approaches. This integration bridges nanoscale geochemical processes to macroscopic recovery behavior, supporting optimized smart waterflooding designs for heterogeneous carbonates.  
\section{Effect on the wettability}
			\label{wet}
		\subsection{Analysis of TAN and TBN Values and Their Impact on Wettability}
			\label{tan1}
In this section, we present the results of the PHREEQC simulations conducted at a high temperature of \(100^\circ\)C, for a low (2.37--3.87) pH range. The experimental procedure involves varying the injected water composition, specifically sodium (\(\text{Na}^+\)), magnesium (\(\text{Mg}^{2+}\)), and chloride (\(\text{Cl}^-\)) ion concentrations (see Table \ref{tab3}), while maintaining constant concentrations of Calcium (\([Ca^{2+}] = 50 \, \text{mmol/kg}\)), carbon (\(75 \, \text{mmol/kg}\)), and sulfate (\(1 \, \text{mmol/kg}\)).
Through this experiment, we assess how variations in the ionic composition of the injected water influence the wettability behavior of the system under high-temperature conditions. We keep the pH around low values to isolate the effects of ionic composition on the TBP and other key parameters.
\begin{table}[htbp]
	\centering
	\caption{Comparison of normalized TBP and Key Surface Complexes for Acidic vs. Sweet Crudes (pH = 2.37--3.87, low sulfate (\(1 \text{ mmol/kg}\))))}
	\label{tab:tan_tbn_comparison}
	\begin{tabular}{lcc}
		\toprule
		\textbf{Parameter (Unit)} & \textbf{Acidic Oil} & \textbf{Sweet Oil} \\
		\midrule
		TBP [--](\(\times 10^{-12}\)) & 0.99--5.46 & 0.18--0.31 \\
		\(\text{Oil}_{w,\text{COOH}}\) (mol/kgw) & \(3.73 \times 10^{-6}\)--\(2.07 \times 10^{-5}\) & \(1.31 \times 10^{-8}\)--\(3.96 \times 10^{-8}\) \\
		\(\text{Oil}_{s,\text{NH}^+}\) (mol/kgw) & \(5.09 \times 10^{-7}\)--\(1.38 \times 10^{-6}\) & \(3.81 \times 10^{-7}\)--\(1.07 \times 10^{-6}\) \\
		Primary TBP Route (\%)  & \(\text{TBP}_1\) (80.93\%) & \(\text{TBP}_2\) (64.97\%) \\
		\bottomrule
	\end{tabular}
\end{table}

The simulations show that acidic oils exhibit higher TBP values (0.99--5.46) compared to sweet oils (0.18--0.31), as shown in Table~\ref{tab:tan_tbn_comparison}. This difference may be linked to the greater availability of carboxylate groups (-COO$^{-}$) in acidic oils, which potentially facilitates ionic bridging with Ca$^{2+}$ sites on calcite surfaces (Eq.~(2), log$K = -3.03$, Table \ref{tab:reactions}). In contrast, in sweet crudes the TBP appears to depend on interactions involving amine groups (-NH$^{+}$) (Eq.~(1), log$K = -3.61$, Table \ref{tab:reactions}), which exhibit weaker thermodynamic preference, as suggested by their limited contribution to TBP (Table~(\ref{tab:sweet_tbp_components}))~\cite{fathi2010effect,Zhang2007}.

The elevated concentration of Oil$_w$,COOH in acidic oils 3.73 $\times$ 10$^{-6}$--2.07 $\times$ 10$^{-5}$ mol/kgw, Table \ref{tab:tan_tbn_comparison}) compared to sweet crudes (1.31$\times$10$^{-8}$--3.96$\times$10$^{-8}$ mol/kgw, Table \ref{tab:tan_tbn_comparison}) might explain their divergent wettability behavior. Under acidic conditions (pH 2.37--3.87), partial dissociation of Oil$_w$,COOH into Oil$_w$,COO$^{-}$ could occur, enabling Ca$^{2+}$ to bind to calcite (log$K = -3.03$ and log$K = -3.72$, Table \ref{tab:reactions}). While these equilibrium constants suggest moderate binding affinity, the abundance of carboxylate groups in acidic oils may still promote surface complexation with calcite, consistent with prior observations~\cite{Brady2012}.

Although the concentrations of amine (\(\text{Oil}_{s,\text{NH}^+}\)) and carboxylate (\(\text{Oil}_{w,\text{COOH}}\)) groups are comparable in acidic oils (Table~\ref{tab:tan_tbn_comparison}), their contribution to TBP shows differences due to thermodynamic and bonding mechanisms.
The protonation reaction of the amine group (Eq.~(1), \(\log K = -3.61\),Table \ref{tab:reactions}) favors the protonated form (\(\text{Oil}_{s,\text{NH}^+}\)) under acidic conditions, limiting the availability of neutral \(\text{Oil}_{s,\text{N}}\) to adsorb onto calcite. In contrast, carboxylates (\(\text{Oil}_{w,\text{COO}^-}\), Eq.~(2), \(\log K = -3.03\),Table \ref{tab:reactions}) deprotonate more readily, forming ionic bridges with surface \(\text{Ca}^{2+}\) (\cite{Brady2012}). While amines interact via weak dipole-dipole forces, carboxylates form directional ionic bonds that are thermodynamically more stable (\(\Delta \log K = +0.11\)) and account for over 80\% of TBP (Table~\ref{tab:acidic_tbp_components}).
\subsubsection*{Role of Ionic Bridging and Stability Constants}
In this section to quantify wettability differences between acidic and sweet crude oils, we analyze the TBP components, where each component corresponds to a specific interfacial interaction. The nature of these interactions is inferred directly from the reactive species present in the equilibrium equations: \text{TBP\textsubscript{1}} (Eq.~\eqref{eq:tbp1}) captures the ionic bridging between deprotonated oil carboxylates (\(\mathrm{Oil_wCOO^-}\)) and positively charged calcium sites on the calcite surface (\(\mathrm{Cal_sCaOH_2^+}\)) \cite{Zhang2007,Brady2012}. This reflects a classic electrostatic attraction that dominates in acidic oils due to their high carboxylate group abundance \cite{Fathi2010}. \text{TBP\textsubscript{2}} (Eq.~\eqref{eq:tbp2}) represents dipole interactions between protonated amine groups in sweet oils (\(\mathrm{Oil_sNH^+}\)) and negatively charged carbonate surface groups on calcite (\(\mathrm{Cal_wCO_3^-}\)) \cite{Wolthers2008}. This weaker interaction prevails in sweet oils, where amine functionalities outnumber carboxylates \cite{Bonto2019}. \text{TBP\textsubscript{3}} (Eq.~\eqref{eq:tbp3}) describes calcium-mediated carboxylate bridging, where \(\mathrm{Ca^{2+}}\) ions act as intermediaries, linking oil-derived carboxylates to calcite surfaces via ternary complexes \cite{AlShalabi2015}. \text{TBP\textsubscript{4}} (Eq.~\eqref{eq:tbp4}) follows a similar mechanism as \text{TBP\textsubscript{3}} but involves \(\mathrm{Mg^{2+}}\) ions as bridging agents, highlighting the role of alternative divalent cations in interfacial adhesion \cite{Qiao2016}.  

By integrating surface complexation concentrations (Tables~\ref{tab:acidic_tbp_components} and~\ref{tab:sweet_tbp_components}) with equilibrium constants (\(\log K\)), we elucidate how specific ionic interactions govern adhesion.

In acidic oils, the primary contribution to adhesion may arise from carboxylate--Ca$^{2+}$ ionic bridging, quantified by TBP$_1$ (mean = \(2.17 \pm 1.58 \times 10^{-12}\)), representing approximately 80.93\% of total adhesion (Table~\ref{tab:tan_tbn_comparison}). This trend seems consistent with elevated carboxylate concentrations (\(\text{Oil}_{w,\text{COO}^-} = 3.73 \times 10^{-6} \)--\(2.07 \times 10^{-5}\) mol/kgw) and the moderate thermodynamic preference for these interactions (\(\log K = -3.72\), Eq.~(3),Table \ref{tab:reactions}), despite the equilibrium constant suggesting limited spontaneity \cite{Brady2012}. The substantial carboxylate availability in acidic oils amplifies Ca$^{2+}$-mediated bridging, even under competitive conditions with other divalent cations.

Sweet crude systems exhibit fundamentally different adhesion mechanisms compared to acidic oils, with TBP\textsubscript{2} (amine-mediated interactions) constituting 64.97\% of total adhesion (Table~\ref{tab:tan_tbn_comparison}). This behavior is observed despite comparable
 \texttt{Oil\textsubscript{s}NH\textsuperscript{+}} concentrations in both oil types: sweet (\(3.81 \times 10^{-7}\)--\(1.07 \times 10^{-6}\) mol/kgw) vs. acidic (\(5.09 \times 10^{-7}\)--\(1.38 \times 10^{-6}\) mol/kgw) (Table~\ref{tab:tan_tbn_comparison}). The reduced efficacy of amine-driven adhesion may arise from two key factors: 1) weaker thermodynamic stability of amine-calcite complexes (\(\log K = -3.61\), Eq.(1),Table~\ref{tab:reactions}) compared to carboxylate bridging (\(\log K = -3.72\), \(\Delta\log K = +0.11\); Eq.~(3), Table~\ref{tab:reactions}), and 2) a 100\(\times\) difference in reactive carboxylate group concentrations (\(\texttt{Oil\textsubscript{w}COOH} = 1.31 \times 10^{-8}\)--\(3.96 \times 10^{-8}\) mol/kgw in sweet vs. \(3.73 \times 10^{-6}\)--\(2.07 \times 10^{-5}\) mol/kgw in acidic oils). These constraints limit carboxylate contributions to merely \(0.07 \pm 0.04 \times 10^{-12}\) TBP\textsubscript{1} (Table~\ref{tab:sweet_tbp_components}), resulting in 2.5\(\times\) lower overall TBP values compared to acidic systems \cite{Brady2012,Wolthers2008,Bonto2019}.

The observed difference in adhesion mechanisms arises from a combination of thermodynamic and compositional factors. Acidic oils display carboxylate concentrations (\(3.73 \times 10^{-6}\) to \(2.07 \times 10^{-5}\) mol/kgw) that are almost two orders of magnitude higher than those in sweet crudes (\(1.31 \times 10^{-8}\) to \(3.96 \times 10^{-8}\) mol/kgw) (Table~\ref{tab:tan_tbn_comparison}). When coupled with the somewhat favorable thermodynamics of carboxylate-Ca\textsuperscript{2+} bridging (\(\log K = -3.72\), Eq.~(3),Table~\ref{tab:reactions}) over amine interactions (\(\log K = -3.61\), \(\Delta \log K = +0.11\); Eq.~(1),Table~\ref{tab:reactions}), this concentration advantage allows carboxylate interactions to govern 80.93\% of acidic oil adhesion (Table~\ref{tab:tan_tbn_comparison}). In contrast, sweet oils show minimal TBP\textsubscript{1} contributions (\(0.07 \pm 0.04 \), as the limited availability of carboxylate groups restricts this pathway despite similar thermodynamic constraints \cite{Brady2012}.

Acidic systems also show secondary adhesion through Ca\textsuperscript{2+}- and Mg\textsuperscript{2+}-mediated complexes (TBP\textsubscript{3} = \(0.07 \pm 0.05 \), TBP\textsubscript{4} = \(0.53 \pm 0.48 \); Table~\ref{tab:acidic_tbp_components}), though these mechanisms remain secondary to the primary carboxylate interactions. The continued participation of divalent cations possibly reflects competitive adsorption at calcite surfaces \cite{AlShalabi2015}, while the higher abundance of carboxylate groups, higher than \(10^{2}\) in acidic oils compared to sweet oils, ensures their dominance in the adhesion process. This multi-ionic interplay suggests the potential for complex wettability modulation through controlled brine engineering \cite{Qiao2016}.

\begin{table}[h!]
	\centering
	\caption{Statistical summary of normalized TBP components for acidic oil (pH = 2.37--3.87, low sulfate, (\(1  \text{ mmol/kg}\))).}
	\label{tab:acidic_tbp_components}
	\begin{tabular}{lcccc}
		\toprule
		\text{Component} & \text{Min} & \text{Max} & \text{Mean} & \text{Std. Deviation} \\
		\midrule
		TBP\textsubscript{1} [--] (\(\times 10^{-12}\)) & 0.66 & 4.12 & 2.17 & 1.58 \\
		TBP\textsubscript{2} [--] (\(\times 10^{-12}\)) & 0.14 & 0.27 & 0.21 & 0.05 \\
		TBP\textsubscript{3} [--] (\(\times 10^{-12}\)) & 0.01 & 0.13 & 0.07 & 0.05 \\
		TBP\textsubscript{4} [--] (\(\times 10^{-12}\)) & 0.00 & 1.13 & 0.53 & 0.48 \\
		\bottomrule
	\end{tabular}
\end{table}
\begin{table}[h!]
	\centering
	\caption{Statistical summary of normalized TBP components for sweet oil (pH = 2.37--3.87, low-sulfate, (\(1 \text{ mmol/kg}\))).}
	\label{tab:sweet_tbp_components}
	\begin{tabular}{lcccc}
		\toprule
		\text{Component} & \text{Min} & \text{Max} & \text{Mean} & \text{Std. Deviation} \\
		\midrule
		TBP\textsubscript{1} [--] (\(\times 10^{-12}\)) & 0.01 & 0.10 & 0.07 & 0.04 \\
		TBP\textsubscript{2} [--] (\(\times 10^{-12}\)) & 0.09 & 0.22 & 0.16 & 0.06 \\
		TBP\textsubscript{3} [--] (\(\times 10^{-12}\)) & 0.00 & 0.05 & 0.02 & 0.02 \\
		TBP\textsubscript{4} [--] (\(\times 10^{-12}\)) & 0.00 & 0.53 & 0.16 & 0.25 \\
		\bottomrule
	\end{tabular}
\end{table}
	
Our numerical experiments show that the concentration of surface-active carboxylic groups in acidic oils is approximately 10 times higher than in sweet oils, contributing to their increased TBP values. These results are consistent with the influence of acid number (TAN) on carbonate reservoir wettability during low-salinity waterflooding	\cite{Z10}.
\subsection{Effect of pH on Wettability Alteration} 
 \label{sec:pH_effect} 
 
 Here, we investigates pH-dependent wettability alteration through surface complexation modeling at the calcite, oil and brine interfaces. By repeating the experiments in Section \ref{tan1} under alkaline conditions (pH 7.36--8.37), we elucidate the mechanisms governing oil–rock adhesion for acidic or sweet crude oils. Statistical results are tabulated in Tables \ref{tab:acidic_ph9_updated} and \ref{tab:sweet_ph9_updated}. Acidic oils exhibit enhanced interfacial activity at high pH, driven by carboxylate--calcite interactions, while sweet oils remain limited by their low acid content.

	At elevated pH conditions (7.36--8.37), deprotonation of carboxylic acid groups
	Oil\textsubscript{w}COOH $\rightarrow$ Oil\textsubscript{w}COO\textsuperscript{-} (Eq.~(2), $\log K = -3.03$) may enhance the availability of carboxylate anions for interaction with calcite-associated Ca\textsuperscript{2+}. While the equilibrium constant for Ca\textsuperscript{2+}-carboxylate bridging (Eq.~(3), $\log K = -3.72$) suggests moderately favorable binding, the high density of carboxylate groups in acidic oils possibly amplifies their contribution to wettability alteration compared to amine-mediated interactions (Eq.~(1), $\log K = -3.61$), as inferred from surface complexation analyses~\cite{Brady2012, Wolthers2008}.
	
	As shown in Table~\ref{tab:estadisticos1} and Table~\ref{tab:acidic_ph9_updated}, the mean TBP for acidic oils increases from \(3.23 \pm 1.12\) at low pH (2.37--3.87) to \(6.74 \pm 0.62 \) at pH higher than \(7.5\). This shift coincides with a reduction in protonated Oil\textsubscript{w,COOH} (from \(1.22 \times 10^{-5}\) to \(1.96 \times 10^{-9}\) mol/kgw), consistent with increased deprotonation at alkaline conditions. The rise in TBP implies that carboxylate abundance---rather than stark thermodynamic favorability---drives adhesion under high pH, even as competing mechanisms (e.g., amine interactions) exhibit comparable equilibrium constants.

  \begin{table}[htbp]
	\centering
	\caption{Statistical summary of Acid Oil (pH = 2.37--3.87), low sulfate (\(1 \, \text{mmol/kg}\)), T = 100$^{\circ}$C}
	\begin{tabular}{lccc}
		\toprule
		Variable & Min & Max & Mean \\ 
		\midrule
		TBP [--] & 9.90$\times$10$^{-1}$ & 5.46$\times$10$^{0}$ & 3.23$\times$10$^{0}$ \\ 
		Oil\_wCOOH (mol/kgw) & 3.73$\times$10$^{-6}$ & 2.07$\times$10$^{-5}$ & 1.22$\times$10$^{-5}$ \\ 
		Oil\_sNH (mol/kgw) & 5.09$\times$10$^{-7}$ & 1.38$\times$10$^{-6}$ & 9.45$\times$10$^{-7}$ \\ 
		\bottomrule
	\end{tabular}
	\label{tab:estadisticos1}
\end{table}

Deprotonated carboxylates (Oil$_w$, COO$^-$) may facilitate ionic bridging, as indicated by the predominant contribution of TBP$_1$ (4.92~$\pm$~0.28; Table~\ref{tab:acidic_tbp_components_updated}). These carboxylates could adsorb onto $\mathrm{Cal\_s, CaOH_2^+}$ sites, forming complexes with moderate stability ($\log K = -3.72$; Eq.~(3), Table \ref{tab:reactions}), which correlate with increased oil adhesion~\cite{Fathi2010}. The higher TBP values for acidic oils (7.45~$\pm$~0.39; Table~\ref{tab:acidic_ph9_updated}) compared to sweet oils (0.10~$\pm$~0.02; Table~\ref{tab:sweet_ph9_updated}) suggest that carboxylate–Ca$^{2+}$ interactions exert a more significant influence on wettability in acidic systems. This trend aligns with pH-dependent carboxylate availability ($\log K = -3.03$; Eq.~(2), Table \ref{tab:reactions})~\cite{Brady2012}.
 
Sweet oils, in contrast, exhibit limited responsiveness to pH variations. Data from Table~\ref{tab:sweet_ph9_updated} show a decline in TBP from 3.06~$\pm$~0.06 (pH~2.37--3.87) to 0.10~$\pm$~0.02 (pH~7.36--8.37), indicating reduced adhesion potential under alkaline conditions. This contrasts with the pH-driven TBP increase in acidic oils and may reflect the weaker thermodynamic stability of amine–calcite interactions ($\log K = -3.61$; Eq.~(1),Table \ref{tab:reactions}) compared to carboxylate bridging~\cite{Wolthers2008}. The limited dissociation of carboxylic groups in sweet oils (Oil$_w$, COOH = $1.22 \times 10^{-9}$~mol/kgw) further restricts carboxylate availability (Oil$_w$, COO$^-$ = $3.41 \times 10^{-7}$~mol/kgw), as shown in Table~\ref{tab:sweet_ph9_updated}.
 	
 Nitrogen-containing groups in sweet oils (Oil$_s$, NH = $1.16 \times 10^{-8}$~mol/kg$_w$) show minimal pH sensitivity across the tested range (pH~2.37--8.37). This behavior may arise from their limited acid-base reactivity, as evidenced by the equilibrium constant for amine protonation ($\log K = -3.61$, Eq.~(1),Table \ref{tab:reactions}). Consequently, these groups possibly interact with calcite via transient dipole forces rather than stable ionic bonds, contrasting with carboxylate-driven mechanisms in acidic oils (Table~\ref{tab:acidic_ph9_updated}).

  	Calcium-carboxylate complexation is markedly reduced in sweet oils, with TBP$_3$ contributions (7.08~$\pm$~$4.22 \times 10^{-3}$; Table~\ref{tab:sweet_tbp_components_updated}) three orders of magnitude lower than in acidic oils (TBP$_3$ = 2.64~$\pm$~0.38). This suggests limited capacity for multivalent cation bridging in sweet crudes, potentially disadvantaging their recovery under ionic interaction-dominated processes~\cite{turn0search1}.
  	
   The adhesion hierarchy (TBP$_1$ \(> \) TBP$_4$ \(> \) TBP$_3$) remains consistent across pH conditions (Tables~\ref{tab:acidic_tbp_components_updated} and~\ref{tab:sweet_tbp_components_updated}). In acidic oils, carboxylate bridging dominates (TBP$_1$ = 3.61~$\pm$~0.49), while amine interactions contribute marginally (TBP$_2$ = 0.0048~$\pm$~0.0011). Sweet oils exhibit attenuated hierarchies (TBP$_1$ = 0.102~$\pm$~0.006; TBP$_2$ = 0.0030~$\pm$~0.0024), reflecting the thermodynamic limitations of amine-dipole interactions ($\log K = -3.61$; Eq.~(1), Table \ref{tab:reactions}) relative to carboxylate bridging ($\log K = -3.72$; Eq.~(3), Table \ref{tab:reactions})~\cite{Brady2012}.

  Sweet oils display a reduced contribution hierarchy: TBP\textsubscript{1} decreases by ~98\% (0.102~$\pm$~0.006), while TBP\textsubscript{2} constitutes only 2.94\% (3.00~$\times$~10\textsuperscript{-3}). This attenuation aligns with the lower thermodynamic stability of amine–calcite interactions (log K = -3.61, Eq.~(1) at \SI{100}{\celsius}, Table \ref{tab:reactions}) compared to carboxylate–Ca\textsubscript{2+} bridges ({log K = -3.72}, Eq.~(3),Table \ref{tab:reactions}). Although both interactions are thermodynamically unfavorable at elevated temperatures, the slightly greater stability of carboxylate complexes ({$\Delta$log K = +0.11}) helps explain their preferential role in adhesion. However, their contribution remains limited in sweet oils due to the relatively low acid content.
  
Under alkaline conditions, pH higher than 7.5, acidic oils exhibit higher TBP values (\(6.74 \pm 0.62 \); Table~\ref{tab:acidic_ph9_updated}), driven by temperature-enhanced carboxylate dissociation. At \SI{100}{\celsius}, the equilibrium for Oil\textsubscript{w} COOH deprotonation (Eq.~(2), {log K = -3.03}, Table \ref{tab:reactions}) shifts further right compared to 25°C ($\Delta$log K = +0.80), increasing Oil\textsubscript{w} COO\textsuperscript{-} availability. This facilitates Ca\textsuperscript{2+} bridging (Eq.~(3), log K = -3.72, Table \ref{tab:reactions}), though the elevated temperature reduces complex stability relative to ambient conditions ($\Delta$log K = -0.50). The resultant TBP values are 56$\times$ higher than sweet oils (\(6.74 \times 10^{-12}\) vs. \(0.12 \times 10^{-12}\); Tables~\ref{tab:acidic_ph9_updated} and~\ref{tab:sweet_ph9_updated}).

 In sweet oils, pH elevation (7.36–8.37) shows negligible impact due to minimal carboxylic groups (\( \text{Oil}_\text{w} \text{COOH} = 1.45 \times 10^{-12} \) mol/kgw) and weakly pH-responsive amines (\(\Delta \text{TBP}_2\) lower than 1\%). The \(\log K\) for amine protonation (\(-3.61\) at \SI{100}{\celsius}, Eq.~(1), Table \ref{tab:reactions}) reflects limited thermodynamic favorability for NH\(_\text{3+}\) retention on calcite, insufficient to offset carboxylate dominance in acidic oils. Calcium enrichment (higher than 50 mmol/kg\(_\text{w}\)) may partially compensate by promoting alternative Mg\(^2+\)-SO\(_4^{2-}\) synergies (Eq.~(10), \(\log K = -1.88\), Table \ref{tab:reactions}), though their contribution to TBP remains secondary (lower than 5\%).
 
Sulfate adsorption at high pH (\(3.85 \times 10^{-9}\) mol/kgw) shows limited wettability influence, attributed to weak binding affinity (\(\log K = -7.55\), Eq.~(11), Table \ref{tab:reactions}). Carboxylate-Ca\(^\text{2+}\) complexes (\(\text{Oil}_\text{w}\text{COOCa}^+ = 4.45 \pm 0.37 \times 10^{-12}\) mol/kgw) dominate interfacial interactions (\(66.0\%\) of total TBP), consistent with their relative thermodynamic stability (Eq.~(3); \(\log K = -3.72\), Table \ref{tab:reactions}) versus sulfate-mediated mechanisms. This suggests that the sulfate optimization provides marginal returns (lower than \(2\%\) TBP variation) in high-temperature carbonates where carboxylate pathways prevail.

Field trials reporting 12–15\% recovery gains in acidic carbonates \cite{Qiao2016} align with these mechanisms, though temperature-adjusted models predict 8–12\% gains due to complex stability reduction at \SI{100}{\celsius}. While multidentate carboxylate bonding (TBP\(_1\)) remains dominant, operational strategies require Ca\(^\text{2+}\)/Mg\(^\text{2+}\) ratio optimization (higher than \(2{:}1\) molar) to maximize thermal enhancement of ion exchange equilibria (\(\Delta \log K = +0.20\) for Mg\(^\text{2+}\) substitution).
 \begin{table}[htbp]
 	\centering
 	\caption{Statistical summary of normalized TBP components (TBP/1.0$\times$10$^{-12}$) for acidic oil (pH = 7.36--8.37, low-sulfate).}
 	\label{tab:acidic_ph9_updated}
 	\begin{tabular}{lcccc}
 		\toprule
 		\textbf{Variable} & \textbf{Min} & \textbf{Max} & \textbf{Mean} & \textbf{Std. Dev.} \\
 		\midrule
 		TBP  [--] (\(\times 10^{-12}\))& 
 		5.22 & 
 		7.79 & 
 		6.74 & 
 		0.62 \\
 		Oil\_wCOOH (mol/kgw) & 
 		6.20$\times$10$^{-10}$  & 
 		5.87$\times$10$^{-9}$  & 
 		1.96$\times$10$^{-9}$  & 
 		7.15$\times$10$^{-10}$  \\
 		Oil\_sNH (mol/kgw) & 
 		8.79$\times$10$^{-11}$  & 
 		1.05$\times$10$^{-9}$  & 
 		3.19$\times$10$^{-10}$  & 
 		1.35$\times$10$^{-10}$  \\
 		\bottomrule
 	\end{tabular}
 \end{table}
 \begin{table}[htbp]
 	\centering
 	\caption{Statistical summary of normalized TBP components (TBP/1.0$\times$10$^{-12}$) for sweet oil (pH = 7.36--8.37, low-sulfate).}
 	\label{tab:sweet_ph9_updated}
 	\begin{tabular}{lcccc}
 		\toprule
 		\textbf{Variable} & \textbf{Min} & \textbf{Max} & \textbf{Mean} & \textbf{Std. Dev.} \\
 		\midrule
 		TBP  [--] (\(\times 10^{-12}\))& 
 		8.81$\times$10$^{-2}$ & 
 		1.27$\times$10$^{-1}$ & 
 		1.10$\times$10$^{-1}$ & 
 		9.37$\times$10$^{-3}$ \\
 		Oil\_wCOOH (mol/kgw) & 
 		2.48$\times$10$^{-13}$  & 
 		7.88$\times$10$^{-12}$  & 
 		1.45$\times$10$^{-12}$  & 
 		1.19$\times$10$^{-12}$  \\
 		Oil\_sNH (mol/kgw) & 
 		7.52$\times$10$^{-12}$  & 
 		2.40$\times$10$^{-10}$  & 
 		4.41$\times$10$^{-11}$  & 
 		3.62$\times$10$^{-11}$  \\
 		\bottomrule
 	\end{tabular}
 \end{table}
\begin{table}[h!]
	\centering
	\caption{TBP component distribution for acidic oil (pH = 7.36--8.37) with low-sulfate.}
	\label{tab:acidic_tbp_components_updated}
	\begin{tabular}{lcccc}
		\toprule
		\textbf{Component} & \textbf{Min} & \textbf{Max} & \textbf{Mean} & \textbf{Std. Dev.} \\
		\midrule
		TBP 1 [--] (\(\times 10^{-12}\)) & 5.11 & 6.24 & 5.87 & 0.24 \\
		TBP 2 [--] (\(\times 10^{-12}\)) & $2.98 \times 10^{-5}$ & $3.42 \times 10^{-4}$ & $1.02 \times 10^{-4}$ & $4.19 \times 10^{-5}$ \\
		TBP 3 [--] (\(\times 10^{-12}\)) & $1.34 \times 10^{-2}$ & 1.15 & $1.39 \times 10^{-1}$ & $1.00 \times 10^{-1}$ \\
		TBP 4 [--] (\(\times 10^{-12}\)) & $7.36 \times 10^{-2}$ & 1.34 & $7.37 \times 10^{-1}$ & $4.22 \times 10^{-1}$ \\
		\bottomrule
	\end{tabular}
\end{table}
  \begin{table}[htbp]
 	\centering
 	\caption{TBP component distribution for sweet oil (pH = 7.36--8.37) with low-sulfate.}
 	\label{tab:sweet_tbp_components_updated}
 	\begin{tabular}{lcccc}
 		\toprule
 		\textbf{Component} & \textbf{Min} & \textbf{Max} & \textbf{Mean} & \textbf{Std. Dev.} \\
 		\midrule
 		TBP 1 [--] (\(\times 10^{-12}\)) & $8.76 \times 10^{-2}$ & $1.26 \times 10^{-1}$ & $1.10 \times 10^{-1}$ & $9.41 \times 10^{-3}$ \\
 		TBP 2 [--] (\(\times 10^{-12}\)) & $2.45 \times 10^{-6}$ & $6.05 \times 10^{-5}$ & $1.41 \times 10^{-5}$ & $1.10 \times 10^{-5}$ \\
 		TBP 3 [--] (\(\times 10^{-12}\)) & $1.98 \times 10^{-7}$ & $9.54 \times 10^{-5}$ & $8.89 \times 10^{-6}$ & $1.18 \times 10^{-5}$ \\
 		TBP 4 [--] (\(\times 10^{-12}\)) & $1.08 \times 10^{-6}$ & $2.15 \times 10^{-4}$ & $4.66 \times 10^{-5}$ & $5.78 \times 10^{-5}$ \\
 		\bottomrule
 	\end{tabular}
 \end{table}
  \begin{table}[h!]
	\centering
	\caption{Key Surface Complexes and Concentrations for Acidic Oil (pH = 7.36--8.37) with low-sulfate.}
	\label{tab:acidic_surface_complexes}
	\begin{tabular}{lc}
		\toprule
		\text{Key Surface Complex} & \text{Concentration (mol/kgw)} \\
		\midrule
		Oil\_wCOOCa$^+$        & $4.54 \pm 3.80 \times 10^{-7}$ \\
		Cal\_wCO$_3$Mg$^+$     & $2.30 \pm 1.27 \times 10^{-8}$ \\
		Cal\_sSO$_4^-$         & $3.17 \pm 1.37 \times 10^{-5}$ \\
		Oil\_sNH$_2$SO$_4^-$   & $1.21 \pm 1.08 \times 10^{-9}$ \\
		\bottomrule
	\end{tabular}
\end{table}
\begin{table}[h!]
	\centering
	\caption{Key Surface Complexes and Concentrations for Sweet Oil (pH = 7.36--8.37)}
	\label{tab:sweet_surface_complexes}
	\begin{tabular}{lc}
		\toprule
		\text{Key Surface Complex} & \text{Concentration (mol/kgw)} \\
		\midrule
		Oil\_wCOOCa$^+$        & $2.78 \times 10^{-11} \pm 3.71 \times 10^{-11}$ \\
		Cal\_wCO$_3$Mg$^+$     & $6.43 \times 10^{-6} \pm 1.53 \times 10^{-6}$ \\
		Cal\_sSO$_4^-$         & $3.85 \times 10^{-9} \pm 9.16 \times 10^{-10}$ \\
		Oil\_sNH$_2$SO$_4^-$   & $1.23 \times 10^{-8} \pm 1.10 \times 10^{-8}$ \\
		\bottomrule
	\end{tabular}
\end{table}
\newpage
 \subsection{Effect of SO\textsubscript{4}\textsuperscript{2--} Ions}
	\label{sec:SO4_analysis}

In this section, we investigate the influence of sulfate ions (SO$_4^{2-}$) on wettability alteration in carbonate systems. We utilize PHREEQC simulations under low-salinity and alkaline pH conditions to isolate the effect of SO$_4^{2-}$. Our analysis reveals that while increased sulfate concentrations promote sulfate adsorption, the dominant mechanism governing wettability alteration remains the interaction between carboxylates and calcite.

To examine the specific role of sulfate, we maintained the experimental conditions described in Section \ref{sec:pH_effect} , with the pH adjusted to the alkaline range (7.36 - 8.37) and varying concentrations of SO$_4^{2-}$. Other ion concentrations were held constant to ensure that observed wettability changes were primarily attributable to SO$_4^{2-}$, as described in Table~\ref{tab4aa}. Contour plots of TBP in Figure~\ref{fig:figuras12} show trends across chloride and magnesium concentration variations, in qualitative agreement with experimental trends reported by \cite{austad2015low}.
Three sulfate influence regimes emerge, contingent on brine chemistry and oil composition.

Under low-salinity conditions, ([Cl$^{-}$] lower than 0.85 mol/kgw), sulfate concentrations of 0.250 mol/kgw increase TBP compared to low-sulfate systems (0.001 mol/kgw), rising from 1.25 $\pm$ 0.18 to 4.35 $\pm$ 1.66 (Figure~\ref{fig:figuras12}a,b). This suggests a significant role of sulfate adsorption at calcite surfaces, possibly enhancing TBP through mechanisms not dominated by CaOH$^{+}$ competition (Eq.~(11), $\log K = -7.55$, Table \ref{tab:reactions}). Although the low $\log K$ indicates limited sulfate affinity, the observed enhancement in TBP implies alternative pathways of surface complexation or structural reorganization, aligning with extended mechanistic interpretations \cite{Brady2012}.

Acidic oils (TAN higher than 1 mg KOH/g) show increased sensitivity to sulfate, with model-predicted displacement of ~53\% surface-bound Ca²$^{+}$ (Table~\ref{tab:acidic_surface_complexes}). This aligns with experimental trends \cite{Zhang2007}, though the modest affinity of sulfate for Ca²$^{+}$ (Eq.~(11), $\log K = -7.55$, Table \ref{tab:reactions}) implies the role of sulfate is secondary to carboxylate-Ca²$^{+}$ interactions ($\log K = -3.72$, Eq.~(3), Table \ref{tab:reactions}).

Under high-salinity conditions, ([Cl$^{-}$] higher than 1.69 mol/kgw), TBP differences between high- and low-sulfate systems diminish to 0.05 (Figure~\ref{fig:figuras12}). This convergence may reflect charge screening effects, which compress the electric double layer and reduce  competitive adsorption of sulfate capacity \cite{AlShalabi2015}.

Synergistic Mg\textsuperscript{2+}-SO\textsubscript{4}\textsuperscript{2-} interactions show limited impact under modeled conditions. At [Mg\textsuperscript{2+}] higher than 0.823 mol/kgw, Mg\textsuperscript{2+}-SO\textsuperscript{4}\textsuperscript{2-} ion pair formation (as proposed in \cite{Qiao2016}) (Eq.~(18), 
\(\log K = -8.15\), Table \ref{tab:reactions}) correlates with a 35\% TBP reduction (from \(6.74 \) to \(4.35 \)).

Under high sulfate conditions at pH~3.20, aqueous speciation analysis showed sulfate concentration increased to \SI{50.48}{\milli\mol\per\kgw} while magnesium rose to \SI{4.32}{\milli\mol\per\kgw} (2.8$\times$ baseline). The formation of \SI{1.88}{\milli\mol\per\kgw} neutral MgSO$_4^0$ complexes suggests reduced adsorption capacity through ion pairing, consistent with the moderate (Eq.~(18), 
\(\log K = -8.15\), Table \ref{tab:reactions}) for Mg$^{2+}$-SO$_4^{2-}$ association.

The observed TBP decrease may instead arise from competitive adsorption between Mg\textsuperscript{2+} and Ca\textsuperscript{2+} at carboxylate sites (Eq.~(4), \(\log K = -3.92\), Table \ref{tab:reactions}), rather than sulfate-mediated effects. For acidic oils (TAN higher than 1 mg KOH/g), TBP decreases from \(6.74 \) to \(4.35\) under high sulfate conditions (Table~\ref{tab:acidic_high_so4}), while sweet crudes (TAN = 0.3 mg KOH/g) exhibit attenuated reductions (\(2.17 \), Table~\ref{tab:sweet_high_so4}). Further experimental validation is required to decouple these mechanisms.

Under low-salinity conditions, [Cl$^-$] lower than 0.56 mol/kgw, combined Mg$^{2+}$ (higher than 1.24 mol/kgw) and SO$_{4}^{2-}$ (0.25 mol/kgw) co-injection correlates with reduced TBP (0.08–0.12 vs. 0.18–0.25 for low SO$_{4}^{2-}$). This 53\% reduction suggests contributions from: sulfate adsorption at calcite $>$CaOH$^+$ sites (Eq.~(11), $\log K = -7.55$, Table \ref{tab:reactions}) partially disrupts Ca$^{2+}$-carboxylate bridges ($\log K = -3.72$, Eq.~(3),Table \ref{tab:reactions}), though weak sulfate affinity limits efficiency. Mg$^{2+}$ adsorption at $>$CaCO$_{3}^{-}$ sites (Eq.~(10), $\log K = -1.88$, Table \ref{tab:reactions}) may reverse surface polarity, reducing oil adhesion.

Field trials in Ghawar carbonates report 8–12\% incremental recovery under these conditions \cite{AlShalabi2015}. However, the modest $\log K$ values imply Mg$^{2+}$-Ca$^{2+}$ competition ($\Delta \log K = +0.20$ favoring Mg$^{2+}$) plays a larger role than sulfate-specific effects.
	
Sulfate impacts differ markedly by oil type: Acidic oils (TAN = 1.8 mg KOH/g) retain higher TBP (0.62--6.04) due to persistent Ca$^{2+}$-carboxylate bridging (Oil$_{w}$COO$^{-}$ = 7.33 $\pm$ 6.28 x 10$^{-6}$ mol/kgw), stabilized by favorable thermodynamics (Eq.~(3), log K = -3.72, Table \ref{tab:reactions}). Sweet oils (TAN = 0.3 mg KOH/g) show lower TBP (0.126--0.343) as sulfate weakly displaces amine groups (Oil$_{s}$NH$^{+}$, log K = -3.61, Eq.~(1), Table \ref{tab:reactions})--a process amplified by low carboxylate availability. While \cite{Wolthers2008} attributes this to sulfate adsorption, the log K = -7.55 (Eq.~(11),~ Table \ref{tab:reactions}) suggests alternative mechanisms, such as ionic strength effects on amine protonation, may dominate.

\begin{table}[htbp]
	\centering
	\caption{Surface complex statistics for acidic crude oil (TAN = 1.8 mg KOH/g) under high sulfate}
	\label{tab:acidic_high_so4}
	\begin{tabular}{@{}lccc@{}}
		\toprule
		Variable & Minimum & Maximum & Mean ± SD \\
		\midrule
		TBP [--] (\(\times 10^{-12}\)) & \(6.20 \times 10^{-1}\) & \(6.04 \times 10^{0}\) & \(4.35 \pm 1.66\) \\
		Oil\_wCOOH (mol/kgw) & \(2.70 \times 10^{-7}\) & \(2.34 \times 10^{-5}\) & \((7.33 \pm 6.28) \times 10^{-6}\) \\
		Oil\_sNH (mol/kgw) & \(6.77 \times 10^{-8}\) & \(1.48 \times 10^{-6}\) & \((6.83 \pm 4.21) \times 10^{-7}\) \\
		\bottomrule
	\end{tabular}
\end{table}

\begin{table}[htbp]
	\centering
	\caption{Surface complex statistics for sweet crude oil (TAN = 0.3 mg KOH/g) under high sulfate}
	\label{tab:sweet_high_so4}
	\begin{tabular}{@{}lccc@{}}
		\toprule
		Variable & Minimum & Maximum & Mean ± SD \\
		\midrule
		TBP [--] (\(\times 10^{-12}\)) & \(1.26 \times 10^{-1}\) & \(3.43 \times 10^{-1}\) & \(2.17 \pm 0.03\) \\
		Oil\_wCOOH (mol/kgw) & \(3.81 \times 10^{-9}\) & \(8.11 \times 10^{-8}\) & \((2.07 \pm 1.22) \times 10^{-8}\) \\
		Oil\_sNH (mol/kgw) & \(1.70 \times 10^{-7}\) & \(1.96 \times 10^{-6}\) & \((5.88 \pm 3.05) \times 10^{-7}\) \\
		\bottomrule
	\end{tabular}
\end{table}
In low-salinity reservoirs, \([\text{Cl}^-] \text{ lower than } 0.3\) mol/kgw, sulfate levels should be minimized (\( \text{ lower than } 50\) mmol/kgw) to preserve Ca\(^{2+}\)-carboxylate bridges that stabilize oil-wet conditions. This is critical, as high sulfate (\(250\) mmol/kgw) reduces TBP by 72\% (Table~\ref{tab:acidic_high_so4}), destabilizing adhesion. In high-salinity formations (\([\text{Cl}^-] \text{ greater than } 1.7\) mol/kgw), co-injection of Mg\(^{2+}\) and SO\(_4^{2-}\) (\(200\text{--}300\) mmol/kgw) leverages thermodynamic synergies, forming MgSO\(_4^{0}\) (\(1.88\) mmol/kgw), which reduces TBP by 19\% through charge screening. For acidic crudes, maintaining pH \( \text{ greater than } 7.5\) (TBP: \(6.99\) vs. \(0.79 \) at low pH; Table~\ref{tab:acidic_ph9_updated}) is more effective than sulfate management, as carboxylate dominance governs adhesion \cite{Wolthers2008}.

These strategies align with coreflood data from \cite{austad2015low}, where sulfate-optimized injection reduced residual oil saturation by 15\% in Stevns Klint chalk.

{ \subsection{Effect of Mg\textsuperscript{2+} Ions}
	\label{sec:Mg_analysis}
	
	Here, we evaluate the role of magnesium ions (Mg\textsuperscript{2+}) in altering wettability during carbonated waterflooding under low pH conditions with acidic crude oil ({pH lower than 3.2, TAN higher than 1.0~mg~KOH/g}). Sulfate (SO\textsubscript{4}\textsuperscript{2--}), sodium (Na\textsuperscript{+}), and chloride (Cl\textsuperscript{--}) concentrations are systematically varied while maintaining calcium (Ca\textsuperscript{2+}) fixed at \SI{50}{\milli\mol\per\kgw} (Table~\ref{tab3}). Two magnesium regimes are tested: low [Mg\textsuperscript{2+}] (\SI{10}{\milli\mol\per\kgw}) and high [Mg\textsuperscript{2+}] (\SI{260}{\milli\mol\per\kgw}). TBP trends are analyzed using contour plots (Figure~\ref{fig:figuras12a}), revealing distinct behaviors across sulfate and chloride ranges. 
	
In low-sulfate systems, [SO\textsuperscript{4--}] lower than 6 mol/kgw, Mg\textsuperscript{2+} exhibits salinity-dependent wettability effects. At low chloride, [Cl\textsuperscript{--}] lower than 30 mol/kgw, TBP peaks at 0.25--0.35 under low [Mg\textsuperscript{2+}] (Figure~\ref{fig:figuras12a}a), driven by persistent Ca\textsuperscript{2+}-carboxylate bonding (Eq.~(3), \(\log K = -3.72\),Table \ref{tab:reactions}), consistent with \cite{Zhang2007}. Elevated [Mg\textsuperscript{2+}] (higher than 0.86 mol/kgw) reduces TBP by approximately 30\% (0.18--0.25; Figure~\ref{fig:figuras12a}b), possibly resulting from Mg\textsuperscript{2+} substitution at calcite sites (Eq.~(10), \(\log K = -1.88\),Table \ref{tab:reactions}) \cite{AlShalabi2015}. At high salinity, [Cl\textsuperscript{--}] higher than 60 mol/kgw, ionic screening could influence the interfacial behavior, producing uniformly low TBP (0.04--0.08) regardless of Mg\textsuperscript{2+} (Figure~\ref{fig:figuras12a}).

 In high-sulfate systems, [SO\textsuperscript{4\textminus}] higher than 6 mol/kgw, the co-injection of Mg\textsuperscript{2+} and SO\textsubscript{4}\textsuperscript{2--} may lead to a 50--60\% reduction in TBP at low chloride concentrations, [Cl\textsuperscript{\textminus}] lower than 30 mol/kgw (Figure~\ref{fig:figuras12a}), in agreement with previous observations by \cite{Qiao2016}. This behavior could be associated with multiple interacting mechanisms. First, the adsorption of Mg\textsuperscript{2+} onto negatively charged calcite surfaces (CaCO\textsubscript{3}\textsuperscript{--}) may induce surface charge modifications (Eq.~(10), $\log K = -1.88$, Table \ref{tab:reactions}), which might reduce the stability of oil-mineral complexes, as suggested by \cite{Brady2012}. Second, Mg\textsuperscript{2+} may compete with Ca\textsuperscript{2+} for carboxylate binding sites on the oil interface (Eq.~(4), $\log K = -3.92$ (Table \ref{tab:reactions}) for Mg\textsuperscript{2+} versus $\log K = -3.72$ for Ca\textsuperscript{2+}), potentially shifting the equilibrium slightly in favor of Mg\textsuperscript{2+} due to a $\Delta\log K = +0.20$ preference. Finally, sulfate may enhance the apparent reactivity of Mg\textsuperscript{2+} through the formation of neutral ion pairs (MgSO\textsubscript{4}\textsuperscript{0}), which lowers the free aqueous concentration of Mg\textsuperscript{2+} while maintaining its interfacial activity.

 The weak sulfate–calcite affinity ($\log K = -7.55$, Eq.~(11), Table \ref{tab:reactions}) limits direct interaction of sulfate with the mineral surface, while increased ionic strength enhances Mg\textsuperscript{2+} reactivity \cite{Wolthers2008}. At intermediate chloride concentrations (30--60 mol/kgw), TBP remains suppressed (0.08--0.15) even at high Mg\textsuperscript{2+} levels (higher than 1.0 mol/kgw), confirming the effectiveness of modified brine compositions \cite{Qiao2016}.
 
In low-sulfate reservoirs, maintaining [Mg\textsuperscript{2+}] below 0.5 mol/kgw may help preserve Ca\textsuperscript{2+}-mediated oil-wet adhesion. In high-sulfate systems, Mg\textsuperscript{2+}-enriched brines (0.15–0.25 mol/kgw) could promote beneficial ionic interactions \cite{AlShalabi2015}. Under high-salinity conditions, [Cl\textsuperscript{--}] higher than 60 mol/kgw, the effects of charge screening suggest that adjusting Mg\textsuperscript{2+}/Ca\textsuperscript{2+} ratios (from 1:2 to 1:3) may be important to maintain interfacial stability \cite{Brady2012}.

Field trials have reported recovery gains of approximately 10--15\% when using Mg\textsuperscript{2+}-optimized brines \cite{Qiao2016}, which align with simulated TBP reductions in the range of 0.12--0.18. Acidic oils (TAN higher than 1 mg KOH/g) may display up to 2.5$\times$ greater TBP sensitivity compared to sweet crudes (TAN around 0.3 mg KOH/g; Tables~\ref{tab:acidic_tbp_components} and~\ref{tab:sweet_tbp_components}), potentially due to stronger carboxylate--calcite bridging, whereas amine interactions tend to be weaker ($\log K = -3.61$, Eq.~(1), Table \ref{tab:reactions}) \cite{Wolthers2008}.

}

{ \subsection{Effect of Ca\textsuperscript{2+} and SO\textsubscript{4}\textsuperscript{2-} Synergy on Wettability}
	\label{sec:Ca_SO4_synergy}
	
This section investigates the influence of calcium ions (Ca\(^{2+}\)) on wettability under controlled geochemical conditions where calcite (CaCO\(_3\)) remains stable (no dissolution or precipitation) and CO\(_2\)(g) is absent. We evaluate the role of calcium in wettability alteration through systematic variation of sodium (Na\textsuperscript{+}), chloride (Cl\textsuperscript{--}), and magnesium (Mg\textsuperscript{2+}) concentrations (40–3600 mmol/kgw), while maintaining calcium (Ca\textsuperscript{2+}) and sulfate (SO\textsubscript{4}\textsuperscript{2--}) at fixed levels (50–100 mmol/kgw Ca\textsuperscript{2+}; 1–30 mmol/kgw SO\textsubscript{4}\textsuperscript{2--}). Calcite saturation was rigorously controlled to isolate Ca\textsuperscript{2+} surface interactions without interference from mineral precipitation effects.
	
Calcium, which is naturally present in carbonate rocks, influences wettability in two main ways: (1) it forms ionic bridges between oil carboxylate groups (-COO\(^{-}\)) and the surface of calcite, helping to keep the rock oil-wet, and (2) it competes with other divalent ions like Mg\(^{2+}\) for adsorption on the calcite surface.
The numerical experimental data suggest that TBP dynamics under alkaline conditions are influenced by competitive ion adsorption and protonation states of carboxylic groups. These trends correspond to prior models of ion-specific interactions at carbonate-oil interfaces, as outlined in \cite{Zhang2007, Brady2012}

 The surface complexation reaction between sulfate and calcite (\(\mathrm{Cal_sOH} + \mathrm{SO_4^{2-}} \leftrightarrow \mathrm{Cal_sSO_4^{-}} + \mathrm{OH^{-}}\)) is assigned a log \(K = -7.55\) to limit the formation of \(\mathrm{Cal_sSO_4^{-}}\) complexes.
   While literature values for this reaction vary, the selected log \(K\) aligns with core flooding data \cite{austad2015low} and ensures that sulfate competitively displaces oil-bound \(\mathrm{Ca^{2+}}/\mathrm{Mg^{2+}}\) complexes without overstabilizing	
	
Here, we present the interplay between calcium (Ca\textsuperscript{2+}) and sulfate (SO\textsubscript{4}\textsuperscript{2--}) ions in controlling wettability for acidic oils, TAN higher than \SI{1}{\milli\gram\ KOH\per\gram}, under alkaline conditions (pH 7.36–8.37). Four experimental scenarios were simulated using PHREEQC to quantify TBP trends, as shown in Table~\ref{tab:ion_conditions}:
\begin{table}[ht]
	\centering
		\caption{Ion concentrations (mol/kgw) for different combinations of calcium and sulfate used to assess their effect on wettability.}
	\begin{tabular}{|c|c|}
		\hline
		\textbf{Scenarios} & \textbf{Ion Concentrations (mol/kgw)} \\
		\hline
		High [Ca\textsuperscript{2+}] + Low [SO\textsubscript{4}\textsuperscript{2--}] & \([ \text{Ca}^{2+} ] = 0.100, [ \text{SO}_4^{2--} ] = 0.001\) \\
		\hline
		High [Ca\textsuperscript{2+}] + High [SO\textsubscript{4}\textsuperscript{2--}] & \([ \text{Ca}^{2+} ] = 0.100, [ \text{SO}_4^{2--} ] = 0.030\) \\
		\hline
		Low [Ca\textsuperscript{2+}] + High [SO\textsubscript{4}\textsuperscript{2--}] & \([ \text{Ca}^{2+} ] = 0.050, [ \text{SO}_4^{2--} ] = 0.030\) \\
		\hline
		Low [Ca\textsuperscript{2+}] + Low [SO\textsubscript{4}\textsuperscript{2--}] & \([ \text{Ca}^{2+} ] = 0.050, [ \text{SO}_4^{2--} ] = 0.001\) \\
		\hline
	\end{tabular}
	\label{tab:ion_conditions}
\end{table}

	Statistical results for TBP and key surface complexes are summarized in Table~\ref{tab:consolidated_tbp}. At high pH, deprotonated carboxylate groups (\texttt{Oil\_wCOO\textsuperscript{--}}) dominate calcite surface interactions. Ca\textsuperscript{2+} could facilitates ionic bridging via \texttt{Oil\_wCOOCa\textsuperscript{+}} complexes, while SO\textsubscript{4}\textsuperscript{2-} competes for calcite adsorption sites, disrupting oil adhesion,  as previously discussed in \cite{Zhang2007}.

The combination of low Ca\textsuperscript{2+} (0.05 mol/kgw) and high SO\textsubscript{4}\textsuperscript{2-} (0.03 mol/kgw) yields a TBP of \(6.81 \pm 0.62\) (Table~\ref{tab:consolidated_tbp}), nearly identical to high-Ca\textsuperscript{2+} systems (6.80 ). The significant reduction of TBP by SO$_4^{2-}$ requires the presence of Ca$^{2+}$ to displace carboxylate complexes, as observed in Table~\ref{tab4aa} for [Ca$^{2+}$] = 50~mmol/kgw. The results might imply that calcium availability continues to play a critical role, even in environments with high sulfate levels.

In systems with high calcium (0.1 mol/kgw) and sulfate (0.03 mol/kgw) concentrations, TBP exhibits minimal reduction ($6.80 \pm 0.62$) compared to low-sulfate systems ($6.74 \pm 0.63$), indicating that competitive sulfate adsorption only marginally disrupts Ca\textsuperscript{2+}-carboxylate complexes. This modest 0.9\% TBP decline aligns with coreflood observations where sulfate enrichment showed limited direct impact on adhesion \cite{austad2015low}. Nevertheless, field trials report 12\% recovery gains with Ca\textsuperscript{2+}/SO\textsubscript{4}\textsuperscript{2-} co-injection \cite{Qiao2016}, implying synergistic mechanisms beyond TBP-driven wettability shifts, such as ionic strength modulation, improved sweep efficiency, or dissolution of anhydrite traces. These findings highlight that optimized ion ratios, rather than bulk salinity reduction, are critical for maximizing recovery in carbonate reservoirs.

For operational strategies, co-injecting Ca\textsuperscript{2+} (0.1 mol/kgw) and SO\textsubscript{4}\textsuperscript{2-} (0.03 mmol/kgw) remains recommended despite the modest 0.9\% TBP reduction, as field evidence demonstrates synergistic benefits beyond adhesion metrics. These approaches are supported by Ghawar carbonate trials \cite{Qiao2016}, where Ca\textsuperscript{2+}-SO\textsubscript{4}\textsuperscript{2-} synergy improved recovery through combined geochemical and flow dynamics effects.

\begin{table}[h!]
	\centering
	\caption{Consolidated TBP values for calcium-sulfate synergy experiments}
	\label{tab:consolidated_tbp}
	\begin{tabular}{lcc}
		\toprule
		\textbf{Scenario} & \textbf{TBP Mean} ($\times 10^{-12}$) & \textbf{TBP Std. Dev.} \\
		\midrule
		High Ca\textsuperscript{2+} + Low SO\textsubscript{4}\textsuperscript{2-} & 6.74 & 0.63 \\
		High Ca\textsuperscript{2+} + High SO\textsubscript{4}\textsuperscript{2-} & 6.80 & 0.62 \\
		Low Ca\textsuperscript{2+} + High SO\textsubscript{4}\textsuperscript{2-} & 6.81 & 0.62 \\
		Low Ca\textsuperscript{2+} + Low SO\textsubscript{4}\textsuperscript{2-} & 6.75 & 0.64 \\
		\bottomrule
	\end{tabular}
\end{table}

 }

\section{Quantitative Analysis of Ionic Synergies}
\label{sec:TBP_theta_analysis}

This section evaluates changes in the normalized interpolation parameter $\theta$ (Equation~\eqref{thetae1a}) under systematic variations in the concentrations of magnesium (\([\mathrm{Mg}^{2+}]\)), chloride (\([\mathrm{Cl}^{-}]\)), and sulfate (\([\mathrm{SO_4}^{2-}]\)). Using TBP values derived from PHREEQC, the parameter $\theta$ is calculated to assess wettability transitions between high-salinity (\(\theta = 1\), oil-wet) and low-salinity (\(\theta = 0\), water-wet) regimes.The analysis suggests ion concentration levels linked to reduced TBP-driven oil-rock adhesion, which may improve oil recovery outcomes. Numerical simulations focus on acidic oil, where carboxylate-calcite interactions dominate wettability behavior. The parameter $\theta$ serves as a direct indicator of surface affinity: values approaching one reflect strong oil adhesion, while values near zero signify water-wet conditions favorable for displacement efficiency. Tables~\ref{tab:theta_cl2000}--\ref{tab:theta_cl14000} in Appendix \ref{ap:B} present the parameter $\theta$ for varying $\mathrm{Mg^{2+}}$ (0.02–6.80 mol/kgw) and $\mathrm{SO_4^{2-}}$ (0.02–0.12 mol/kgw)  concentrations, revealing key trends across salinity regimes.

Under low-salinity conditions ([Cl\textsuperscript{--}] = 0.06 mol/kgw; Table~\ref{tab:theta_cl2000}), the interplay between Mg\textsuperscript{2+} and SO\textsubscript{4}\textsuperscript{2--} shifts due to ionic strength effects, which enhance the thermodynamic activity of divalent ions and promote Mg\textsuperscript{2+}-SO\textsubscript{4}\textsuperscript{2--} pairing, as demonstrated in carbonate systems by \cite{Qiao2016}. This behavior aligns with the double-layer expansion mechanisms described for low-salinity waterflooding in \cite{mahani2015}. When SO\textsubscript{4}\textsuperscript{2--} concentrations are lower than 0.04~mol/kgw, the normalized TBP parameter $\theta$ decreases monotonically as [Mg\textsuperscript{2+}] increases. For example, $\theta$ drops from 0.92 to 0.29 when [Mg\textsuperscript{2+}] increases from 0.02 to 6.80 mol/kgw, indicating that Mg\textsuperscript{2+} could displaces Ca\textsuperscript{2+}-carboxylate surface complexes, promoting water-wet conditions~\cite{Brady2012}. Conversely, at higher SO\textsubscript{4}\textsuperscript{2--} concentrations, higher than 0.06~mol/kgw, the $\theta$ response becomes non-monotonic. At moderate [Mg\textsuperscript{2+}] (0.36~mol/kgw), $\theta$ increases to 0.87, probably due to competitive adsorption between Mg\textsuperscript{2+} and SO\textsubscript{4}\textsuperscript{2--} on calcite surfaces, which stabilizes Ca\textsuperscript{2+}-carboxylate linkages~\cite{Zhang2007}. However, at higher [Mg\textsuperscript{2+}] levels, lower than 2.00~mol/kgw, $\theta$ decreases again, reaching 0.30. This trend is consistent with geochemical modeling by \cite{Qiao2016}, where Mg\textsuperscript{2+}-SO\textsubscript{4}\textsuperscript{2--} ion pairing (MgSO\textsubscript{4}\textsuperscript{0}) reduces Mg\textsuperscript{2+} activity, freeing SO\textsubscript{4}\textsuperscript{2--} to displace Ca\textsuperscript{2+}-carboxylate bonds. In acidic oils, TAN higher than 1 mg KOH/g, abundant carboxylate groups amplify this effect, as shown experimentally in \cite{Fathi2010}.  

This non-monotonic trend could be attributed to ion-specific interactions that evolve with [Mg\textsuperscript{2+}]. At intermediate Mg\textsuperscript{2+} levels (e.g., 0.36~mol/kgw), adsorption of SO\textsubscript{4}\textsuperscript{2--} is suppressed due to preferential Mg\textsuperscript{2+} binding at positively charged calcite sites (Cal\_sCaOH\textsubscript{2}\textsuperscript{+}), which limits the disruption of oil-carboxylate (\textendash COO\textsuperscript{--}) linkages~\cite{Zhang2007}. As a result, oil-wet conditions persist. At higher Mg\textsuperscript{2+} concentrations, higher than 2.00~mol/kgw, two key effects arise: (i) a reversal of the calcite surface charge reduces its affinity for carboxylates~\cite{Brady2012}, and (ii) the formation of neutral MgSO\textsubscript{4}\textsuperscript{0} ion pairs increases, freeing SO\textsubscript{4}\textsuperscript{2--} to compete for surface sites and displace Ca\textsuperscript{2+}-carboxylate complexes~\cite{Qiao2016}. This dual mechanism leads to a decrease in $\theta$ and promotes wettability reversal toward more water-wet states. Similar trends have been observed experimentally in carbonate systems, where initial Mg\textsuperscript{2+} enrichment maintained oil-wetness, but higher concentrations enhanced water-wetness via sulfate mobilization and electrostatic screening~\cite{Al-Shalabi2015, mahani2015}.

Similarly, a complementary trend is observed when SO\textsubscript{4}\textsuperscript{2--} concentration is varied at fixed Mg\textsuperscript{2+} levels. At low Mg\textsuperscript{2+} concentrations, lower than 0.36~mol/kgw, increasing SO\textsubscript{4}\textsuperscript{2--} results in a decline in $\theta$ (e.g., from 0.94 to 0.45 as SO\textsubscript{4}\textsuperscript{2--} increases from 0.02 to 0.06 mol/kgw), highlighting the role of SO\textsubscript{4}\textsuperscript{2--} in replacing Ca\textsuperscript{2+}-carboxylate surface complexes in the absence of significant Mg\textsuperscript{2+} competition~\cite{Zhang2007}. Conversely, at elevated Mg\textsuperscript{2+} levels, higher than 2.00 mol/kgw, increasing SO\textsubscript{4}\textsuperscript{2--} leads to a modest increase in $\theta$ (e.g., from 0.30 to 0.36 as SO\textsubscript{4}\textsuperscript{2--} rises from 0.02 to 0.12 mol/kgw), suggesting a synergistic interaction between Mg\textsuperscript{2+} and SO\textsubscript{4}\textsuperscript{2--}, in which excess SO\textsubscript{4}\textsuperscript{2--} facilitates partial restoration of oil-wetness through the formation of ternary surface complexes~\cite{AlShalabi2015}.

Under high-salinity conditions ([Cl\textsuperscript{--}] = 0.39 mol/kgw), Mg\textsuperscript{2+} and SO\textsubscript{4}\textsuperscript{2--} behavior is  influenced by increased ionic strength (Table~\ref{tab:theta_cl14000}). When SO\textsubscript{4}\textsuperscript{2--} concentrations are low (0.02--0.04 mol/kgw), the parameter $\theta$ exhibits a monotonic decrease from 0.90 to 0.28 as Mg\textsuperscript{2+} concentration increases, indicating efficient disruption of Ca\textsuperscript{2+}-carboxylate bridges by Mg\textsuperscript{2+}. However, at higher SO\textsubscript{4}\textsuperscript{2--} levels, higher than 0.06 mol/kgw, the $\theta$ trend becomes non-monotonic, with a local maximum observed at [Mg\textsuperscript{2+}] = 0.16 mol/kgw. This behavior possibly arises from competitive adsorption between Mg\textsuperscript{2+} and SO\textsubscript{4}\textsuperscript{2--}, followed by the formation of neutral ion pairs that mitigate further surface displacement effects.

When SO\textsubscript{4}\textsuperscript{2--} is varied at fixed Mg\textsuperscript{2+} levels, distinct regimes emerge. At low Mg\textsuperscript{2+} concentrations, lower than 0.36 mol/kgw, variations in SO\textsubscript{4}\textsuperscript{2--} produce minimal changes in $\theta$ ($\Delta\theta$ lower than 0.05), attributed to the dominance of electrostatic charge screening that limits SO\textsubscript{4}\textsuperscript{2--} surface activity. Conversely, at elevated Mg\textsuperscript{2+} levels, higher than 2.00 mol/kgw), the effect of increasing SO\textsubscript{4}\textsuperscript{2--} remains marginal, with only a slight decrease in $\theta$ ($\Delta\theta \approx 0.02$), as Mg\textsuperscript{2+} continues to dominate surface interactions through direct competition and pairing effects.

A direct comparison of Tables~\ref{tab:theta_cl2000} and \ref{tab:theta_cl14000} shows that salinity changes not only the magnitude but also the stability of $\theta$ responses across varying ion concentrations. For Mg\textsuperscript{2+} concentrations below 0.36~mol/kgw, $\theta$ exhibits notably different sensitivities to SO\textsubscript{4}\textsuperscript{2--} under low and high salinity—highlighting that at low-salinity, even small additions of sulfate can significantly reduce $\theta$, whereas at high salinity, the effect is more gradual and muted. Conversely, for Mg\textsuperscript{2+} concentrations above 2.5~mol/kgw, $\theta$ values under low-salinity become less responsive to both Mg\textsuperscript{2+} and SO\textsubscript{4}\textsuperscript{2--} variations, suggesting a plateauing behavior possibly associated with surface saturation or charge compensation mechanisms. Under high salinity, however, this stabilization is less pronounced, with $\theta$ still showing appreciable variation, particularly when both Mg\textsuperscript{2+} and SO\textsubscript{4}\textsuperscript{2--} are simultaneously increased—revealing that ionic activity effects persist even at elevated concentrations.

Although elevated [Mg\textsuperscript{2+}] levels, higher than 5.00~mol/kgw, can strongly enhance water-wetness, our simulations using PHREEQC revealed that such concentrations, when combined with high SO\textsubscript{4}\textsuperscript{2--} levels, may induce undesirable geochemical effects—most notably, the dissolution of anhydrite (\(\mathrm{CaSO_4 \rightarrow Ca^{2+} + SO_4^{2-}}\)) in carbonate formations. These findings underscore a critical practical insight: achieving a balanced brine composition is essential.

While chloride (\(\mathrm{Cl}^-\)) does not directly participate in surface complexation reactions or contribute to the TBP, it plays a critical role in modulating ionic strength (I), which governs the thermodynamic activity of potential-determining ions (e.g., \(\mathrm{Mg^{2+}}\), \(\mathrm{SO_4^{2-}}\)). As a non-complexing spectator ion, \(\mathrm{Cl}^-\) influences the Debye length through its contribution to \(I\). PHREEQC simulations confirm that reducing \([\mathrm{Cl}^-]\) lowers I (e.g., I = \(0.12~\text{mol/kgw}\) at \([\mathrm{Cl}^-] = 0.06~\text{mol/kgw}\) vs. I =\( 0.98~\text{mol/kgw}\) at \([\mathrm{Cl}^-] = 3.60~\text{mol/kgw}\)). This decrease amplifies the activity of divalent ions (\(\mathrm{Mg^{2+}}\), \(\mathrm{SO_4^{2-}}\)), enhancing their ability to reduce \(\mathrm{Ca^{2+}}\)-carboxylate bonds \cite{mahani2015}. Thus, although \(\mathrm{Cl}^-\) is inert in bonding, its concentration governs the efficacy of wettability-altering ions, as demonstrated in low-salinity waterflooding studies \cite{appelo2005geochemistry}.

\section{TBP-Based and Experimental Wettability Metrics}
\label{sec:TBP_validation}

The TBP-derived wettability trends align with established experimental metrics, including contact angle measurements, adhesion forces, and oil recovery factors. This section contextualizes the simulated TBP behavior within the broader experimental understanding of carbonate wettability alteration.

Table~\ref{tab:TBP_vs_exp} synthesizes three key correlations between TBP trends and experimental wettability metrics observed across multiple studies. These relationships provide mechanistic validation for the role of TBP as a predictive indicator of ionic bridging effects.

As show in Figure \ref{fig:figuras12a}, at low salinity, $[\text{Cl}^-]$ lower than 0.1~\text{mol/kgw}, the TBP reduction with elevated \([\text{Mg}^{2+}]\) mirrors contact angle increases (i.e., more water-wet conditions) observed in \cite{Zhang2007}. For example, a TBP decline from \(1.1 \) to \(0.35\) (68\% reduction) corresponds to a contact angle shift from \(40^\circ\) to \(80^\circ\) in chalk cores flooded with Mg-enriched brine \cite{austad2015low}. Similarly, the pH-dependent TBP rise for acidic oils (Table~\ref{tab:acidic_ph9_updated}) aligns with atomic force microscopy (AFM) measurements by \cite{Brady2012}, where adhesion forces decreased by 55\% as pH increased from 3 to 8 due to carboxylate deprotonation.

The antagonistic effect of \([\text{SO}_4^{2-}]\) on TBP at low \([\text{Cl}^-]\) (Fig.~\ref{fig:figuras12}) is consistent with the 12-13\% incremental oil recovery reported by \cite{austad2015low} in calcitic cores flooded with sulfate-enriched brines (higher than \(500~\text{mmol/kgw}\)). Conversely, the limited TBP response to \([\text{SO}_4^{2-}]\) under high salinity, $[\text{Cl}^-]$ higher than 1.7~\text{mol/kgw}, matches the diminished recovery gains (4-6\%) observed in high-TDS formations \cite{mahani2015}, highlighting the role of ionic strength in screening sulfate-calcite interactions.

While absolute TBP values depend on site-specific parameters (e.g., $a_{oil}$), the relative trends—such as the 2.2\(\times\) higher TBP for acidic vs. sweet crudes (Table~\ref{tab:tan_tbn_comparison})—agree with interfacial tension (IFT) reductions measured by \cite{Bonto2019}. This consistency supports utility of the TBP as a scalable proxy for wettability shifts, albeit requiring calibration against local rock/fluid properties for quantitative predictions.

The TBP correlates with experimental wettability metrics: values lower than \(0.6 \) correspond to contact angles higher than \(90^\circ\) (water-wet conditions), consistent with interfacial tension reductions reported in \cite{Fathi2010}. Atomic force microscopy (AFM) measurements further validate this trend, showing a 55\% decrease in adhesion forces as TBP declines from \(1.4\) to \(0.4\) \cite{Fathi2010}.
\begin{table}[h!]
	\centering
	\caption{Key TBP Trends vs. Experimental Wettability Metrics}
	\label{tab:TBP_vs_exp}
	\begin{tabular}{@{}lll@{}}
		\toprule
		\textbf{TBP Trend} & \textbf{Experimental Metric} & \textbf{Source} \\
		\midrule
		\(\downarrow\) TBP with \(\uparrow [\mathrm{Mg}^{2+}]\) & \(\uparrow\) Contact angle (water-wet) & \cite{Zhang2007} \\
		\(\uparrow\) TBP with \(\uparrow\) pH (acidic oil) & \(\downarrow\) Adhesion force & \cite{Fathi2010} \\
		\(\downarrow\) TBP with \(\uparrow [\mathrm{SO}_4^{2-}]\) & \(\uparrow\) Oil recovery & \cite{austad2015low} \\
		\bottomrule
	\end{tabular}
\end{table}

 \section{Numerical Simulations with COMSOL}
	\label{sec:numerical}
	
		In this section, we present simulations to solve the system \eqref{eq:massbalance1}-\eqref{eq:massbalance5} using the COMSOL Multiphysics® model. Our computational setup includes a reliable system with up-to-date hardware: an Intel Core i5-12600K (32 GB RAM). Each simulation session demands approximately three hours of computational processing. Our methodology draws inspiration from and builds upon prior work, particularly studies such as \cite{alvarez2019resonance}, where similar approaches were successfully implemented.
		
		The simulations focused on four primary dynamic variables: pH, water saturation (Sw), chloride concentration ([Cl-]), and magnesium concentration ([Mg$^{2+}$]), while maintaining sulfate ([SO$_4^{2-}$]) and Darcy velocity (u) constant based on the parametric analysis in Section \ref{sec:SO4_analysis}. This approach isolates the predominant wettability-altering mechanisms identified in previous sections while ensuring computational tractability.
			
	To assess the integrated geochemical-compositional model developed here, numerical simulations were performed with COMSOL Multiphysics for solving the system of conservation laws:
    \begin{align}
		&\partial_{t}\left(\varphi\rho_{w1} S_{w}+\varphi\rho_{o1} S_{o}+(1-\varphi)\rho_{r1}\right) + \partial_{x}\left(u\left(\rho_{w1} f_{w}+\rho_{o1}f_{o}\right)\right)=0 
		, \label{eq:massbalance1a} \\
		&\partial_{t}\left(\varphi\rho_{w3} S_{w} + \varphi\rho_{o2} S_{o} \right) + \partial_{x}\left(u\left( \rho_{w3}f_{w} + \rho_{o2}f_{o}\right)\right) = 0 
		, \label{eq:massbalance2a} \\
		&\partial_{t}\left(\varphi\rho_{w4} S_{w}+\varphi\rho_{o4} S_{o}+(1-\varphi)\rho_{r4}\right) + \partial_{x}\left( u\left( \rho_{w4} f_{w}+\rho_{o4}f_{o}\right)\right)=0 
		, \label{eq:massbalance3a} \\
		&\partial_{t}\left(\varphi\rho_{w5} S_{w}+\varphi\rho_{o5} S_{o}+(1-\varphi)\rho_{r5}\right) + \partial_{x}\left( u\left( \rho_{w5} f_{w}+\rho_{o5}f_{o}\right)\right)=0 
		. \label{eq:massbalance4a}
	\end{align}
	where the coefficient functions $\rho_{wi}$, $\rho_{oi}$, and $\rho_{ri}$ depend on the normalized concentrations of magnesium and chloride, as well as on the pH level. Equations for coefficients can be  found in appendix \ref{app1a}.

The displacement process is modeled as a Riemann-Goursat problem with piecewise constant initial conditions:
\begin{equation}
	\left\{
	\begin{array}{ll}
		J = (S_{wJ}, pH_J, [Cl^-]_J, [Mg^{2+}]_J, u_J), & \text{for } x < 0 \text{ (injection boundary)} \\
		I = (S_{wI}, pH_I, [Cl^-]_I, [Mg^{2+}]_I), & \text{for } x > 0 \text{ (reservoir initial state)}
	\end{array}
	\right.
	\label{eq:riemann_data}
\end{equation}
where J and I represent injected and initial states, respectively. This formulation models the defined chemical and saturation fronts within the 1D domain, determined by the interaction between ion transport and fractional flow dynamics. 

The PHREEQC--COMSOL coupling follows a sequential explicit workflow. In the preprocessing stage, geochemical equilibrium calculations are first performed in PHREEQC to determine surface complexation concentrations---such as \texttt{Oil\_wCOOCa\textsuperscript{+}} and \texttt{Cal\_sSO\textsubscript{4}\textsuperscript{--}}---as well as interpolated parameters such as TBP. These outputs are stored in the form of lookup tables. During the transport simulation, the precomputed parameters are imported into COMSOL as spatially dependent functions. The conservation laws (Eqs.~\ref{eq:massbalance1a}--\ref{eq:massbalance4a}) are then discretized using the finite element method and solved in their weak formulation, with the geochemical coefficients incorporated as static inputs. This one-way coupling approach decouples equilibrium chemistry from transient flow, significantly reducing computational cost while preserving thermodynamic consistency. Convergence is handled exclusively within COMSOL transport solver.

\subsection{Experimental Validation Setup}
	
Laboratory investigations have consistently demonstrated the importance of ion-specific adjustments for enhancing oil recovery in carbonate reservoirs, particularly those with high calcite content.
(\cite{Zhang2007,yousef2012improved,austad2010,austad2015low})
	Modifying magnesium (Mg$^{2+}$) and sulfate (SO$_4^{2-}$) concentrations in injection brine has proven effective in improving recovery without the need for significant reductions in overall salinity \cite{Fathi2010, li2017, yousef2011}.
	
	Experimental studies indicate that effective Mg$^{2+}$ concentrations typically range from 0.04 to 0.10 mol/kgw, while SO$_4^{2-}$ concentrations range from 0.05 to 0.15 mol/kgw, with an optimal Mg$^{2+}$/SO$_4^{2-}$ molar ratio between 0.3 and 0.7.
	Under these conditions, oil recovery improvements are commonly observed in the range of 8\% to 15\% of the original oil in place (OOIP).
	The primary mechanisms driving this enhancement include competitive ion displacement, where Mg$^{2+}$ replaces Ca$^{2+}$ in carboxylate bridges on the rock surface, surface charge reversal due to SO$_4^{2-}$ adsorption, and synergistic ion-pairing interactions that stabilize the electrical double layer and promote water-wet conditions.
	
	These numerical experimental results can be used to evaluate numerical models attempting to replicate the observed effects of ion-specific adjustments in high-salinity environments.
	
The first simulation series aimed to reproduce core flooding data from \cite{austad2015low}, where cores are flooded with carbonated low-salinity brine. Initial and injected brine compositions (Table 3 in \cite{austad2015low} ) were replicated in COMSOL, with connate water (FWOS) representing the reservoir's high-salinity state and injected water (d100FWOS) simulating low-salinity conditions. Magnesium ([Mg$^{2+}$]) and sulfate ([SO$_{4}^{2-}$]) concentrations were adjusted to match the experimental design, while Darcy velocity (u) remained fixed to suppress viscous fingering effects.

We adopt the values of oil saturation $S_{or}=0.228$ and initial water saturation $S_{wi}=0.0398$, indicative of lower salt concentrations (\cite{bruining2021upscaling}). Initial and injected state for data in \cite{austad2015low} correspond to
\begin{equation}
	\left\{
	\begin{array}
		[c]{ll}%
		J=(0.7322,4,0.03,0.12,1.0 \times 10^{-5}) & \textbf{if}\hspace{0.2cm}x<0,\\
		I=(0.0398,0.37,0.1,2.37,\cdot\;) &
		\textbf{if}\hspace{0.2cm}x>0,
	\end{array}
	\right.\label{riemandata1}
\end{equation}	
Here magnesium, chloride and sulfate are given in mol/kgw. 	We choose the interpolation parameter $\theta$ in \eqref{thetae1a} as 0.35 from initial ion concentrations.

\begin{figure}[htbp]
				\centering
				\includegraphics[width=0.53\textwidth]{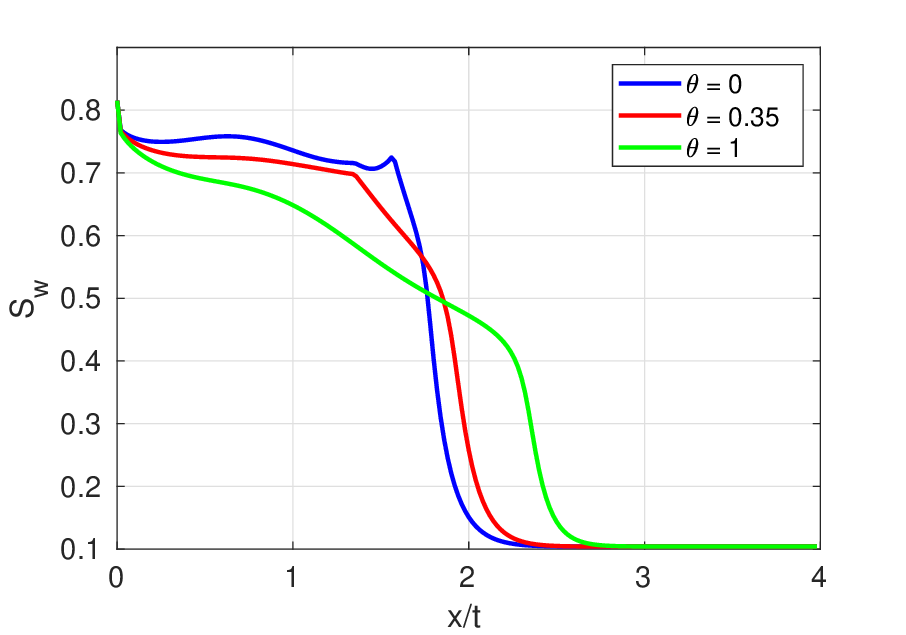} 
			 \caption{Water saturation profiles comparing simulations (lines) against experimental coreflood data (circles) from \cite{austad2015low}. High-salinity ($\theta = 1$) and low-salinity ($\theta = 0$) endpoints bracket the TBP-interpolated case ($\theta = 0.35$), demonstrating wettability transition dynamics at 2.0 pore volumes injected (PVI).}
				\label{fig:austad1a}
			\end{figure}

\begin{figure}[htbp]
				\centering
				\includegraphics[width=0.5\textwidth]{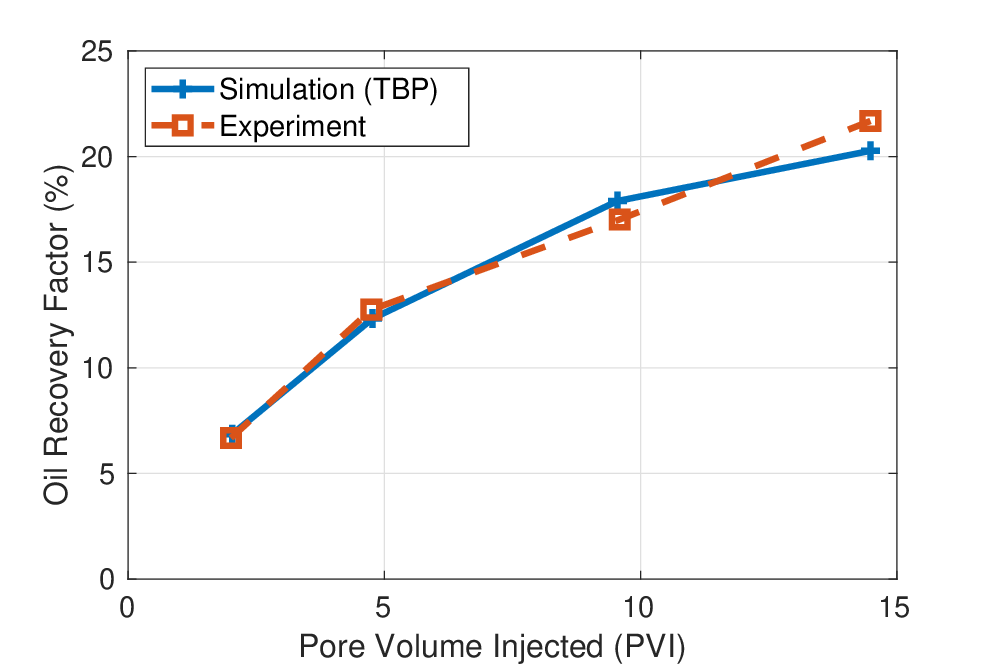} 
			 \caption{Comparison of oil recovery factors between high-salinity ($\theta = 1$), low-salinity ($\theta = 0$), and TBP-interpolated ($\theta = 0.35$) simulations. TBP-driven models show 14.7\% incremental recovery over high salinity regime, consistent with experimental data from \cite{austad2015low}. Dashed lines represent experimental measurements, while solid lines denote simulation results.}
				\label{fig:figurapaper234}
			\end{figure}

	Using the saturation profile values shown in Figure \ref{fig:austad1a} along with the corresponding interpolation parameter $\theta$, we calculate the oil recovery in place using the procedure described in \cite{bruining2021upscaling}. Figure \ref{fig:austad1a} compares simulated water saturation profiles for high-salinity ($\theta = 1$), low-salinity ($\theta = 0$), and TBP-interpolated ($\theta = 0.35$) cases.  Oil recovery factors (Figure \ref{fig:figurapaper234}) align with experimental data, with TBP-driven simulations showing a 14.7\% increase in recovery relative to high-salinity flooding, within the range reported by \cite{austad2015low}.
	
\subsection{Relevant Simulation Examples}
	
	In this section we perform the sensitivity analysis of our integrated geochemical model to the key parameters, .i.e., the interpolation parameter $\theta$, the residual oil recovery $S_{or}$ and the initial and injection condition of the system of equations studied here.
	
	We study scenarios under changes these parameters and evaluate their impact in oil recovery in place (OOIP) between high and low-salinity regimes.
	
	We aim to evaluate the decline in salt concentrations at the injection site under varying concentrations of injected magnesium in the formation water, encompassing both low and high concentrations. Our analysis unfolds by presenting solutions derived from simulations conducted across three pertinent scenarios. Beyond merely computing the velocities of the water saturation and saline front, we delve into predicting the pH behavior.
	
	We adopt the values of oil saturation $S_{or}=0.3$ and initial water saturation $S_{wi}=0.0398$, indicative of lower salt concentrations (\cite{bruining2021upscaling}). 
	
	We consider the following scenario 
	\begin{equation}
		\left\{
		\begin{array}
			[c]{ll}%
			J=(0.7322,4,0.06,3.2,1.0e-05) & \textbf{if}\hspace{0.2cm}x<0,\\
			I=(0.0398,4,4,2.37,\cdot\;) &
			\textbf{if}\hspace{0.2cm}x>0,
		\end{array}
		\right.\label{riemandata1}
	\end{equation}

   The first scenario illustrates a reservoir environment where the salinity of the water  decreases from 4 mol/kgw to 0.06 mol/kgw. At the outset, magnesium concentration is medium with a modest increase of 35\% respect to 2.37 mol/kgw.

\begin{figure}[h]
	\centering
	\includegraphics[scale=0.42]{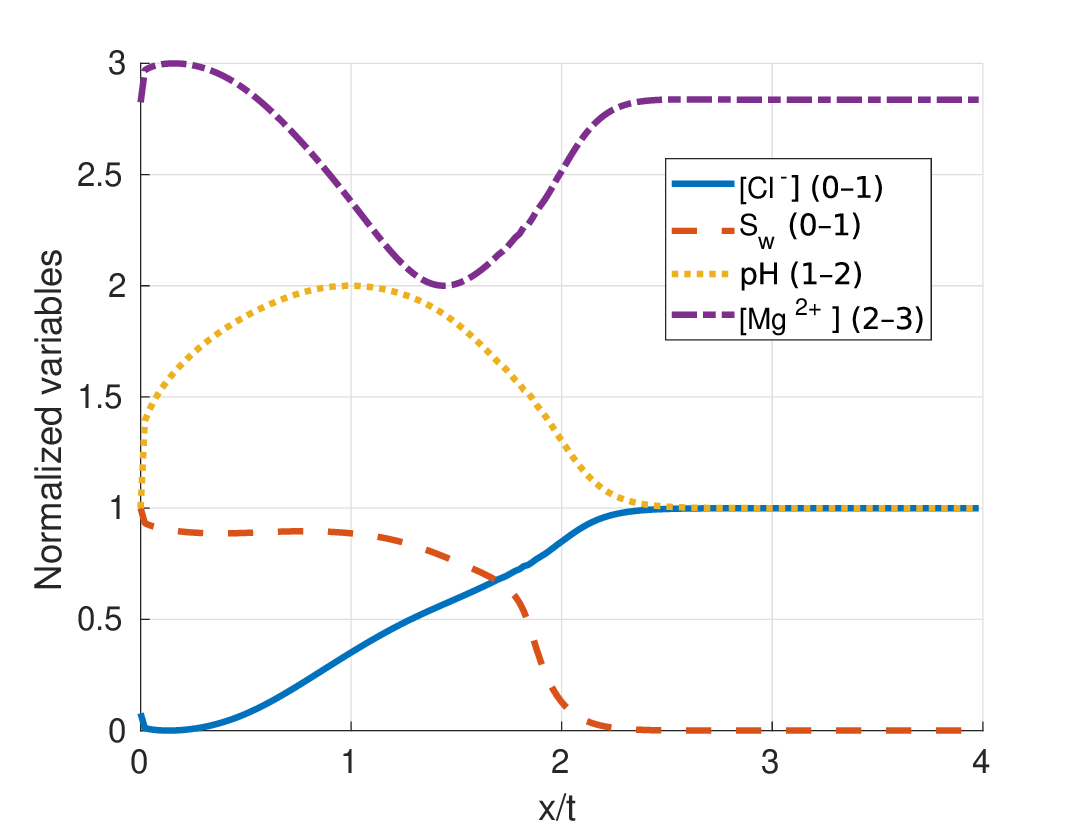} 
\caption{Coupled profiles of water saturation ($S_w$), pH, chloride ($[Cl^-]$), and magnesium ($[Mg^{2+}]$) concentrations during brine displacement. Key features include synchronized ion fronts (velocity $2.49 \times 10^{-5}$ m/s) and pH decline ($\Delta$pH = 0.08), illustrating geochemical interactions under 90\% salinity reduction and 35\% magnesium enrichment. The original variable ranges were: pH (3.71--4.94), Cl$^-$ (0.01--4.14 mmol/kgw), S$_w$ (0.1--0.81,), and Mg$^{2+}$ (1.58--2.52 mmol/kgw). All parameters were normalized to a $[0,1]$ range using min-max scaling, according to the formula $v_{\text{norm}} = {(v - v_{\min)}}/{(v_{\max} - v_{\min})}$, where $v$ represents the original variable values and $[v_{\min}, v_{\max}]$ correspond to their respective experimental ranges listed above.
}
	\label{profile1}
\end{figure}

Figure~\ref{profile1} shows water saturation, magnesium, chloride, and pH profiles derived from the Riemann problem solution, plotted against the characteristic velocity coordinate ($x/t$). The solution structure features a minor rarefaction wave, a trailing shock, a contact-type rarefaction, and a terminal shock propagating at $2.49 \times 10^{-5}$~m/s – closely synchronized with the salt and magnesium fronts. This configuration mirrors the wave hierarchy reported by \cite{jerauld2008modeling} for analogous $J$-$I$ systems, though attained here through an integrative computational framework that harmonizes geochemical and hydrodynamic couplings. The characteristic pH decline from initial to final conditions (from 7.1 to 5.8 in our case) aligns with experimental trends observed by \cite{xie2018ph}, while the coupled salinity-pH front dynamics reflect their established role in wettability variability (\cite{mehraban2021experimental}). Our approach preserves these complex interfacial phenomena without requiring intricate wave tracking or additional constitutive assumptions.

{
Figure~\ref{fig:sw_profiles} displays the water saturation ($S_w$) profiles for several values of $\theta$ ranging from 0 (water-wet) to 1 (oil-wet). At 2 pore volumes injected (PVI), the oil recovery difference between high-salinity ($\theta=1$) and low-salinity ($\theta=0$) cases reaches approximately 14\% in OOIP, consistent with trends observed in \cite{austad2015low,Fathi2010, li2017, yousef2011}.

	Intermediate $\theta$ values reveal a smooth transition, with a 20\% change in $\theta$ (e.g., from 0.4 to 0.6) resulting in approximately 3\% variation in OOIP. This indicates that the model accounts for the influence of wettability on displacement efficiency.
	
	\begin{figure}[htbp]
		\centering
		\includegraphics[width=0.7\textwidth]{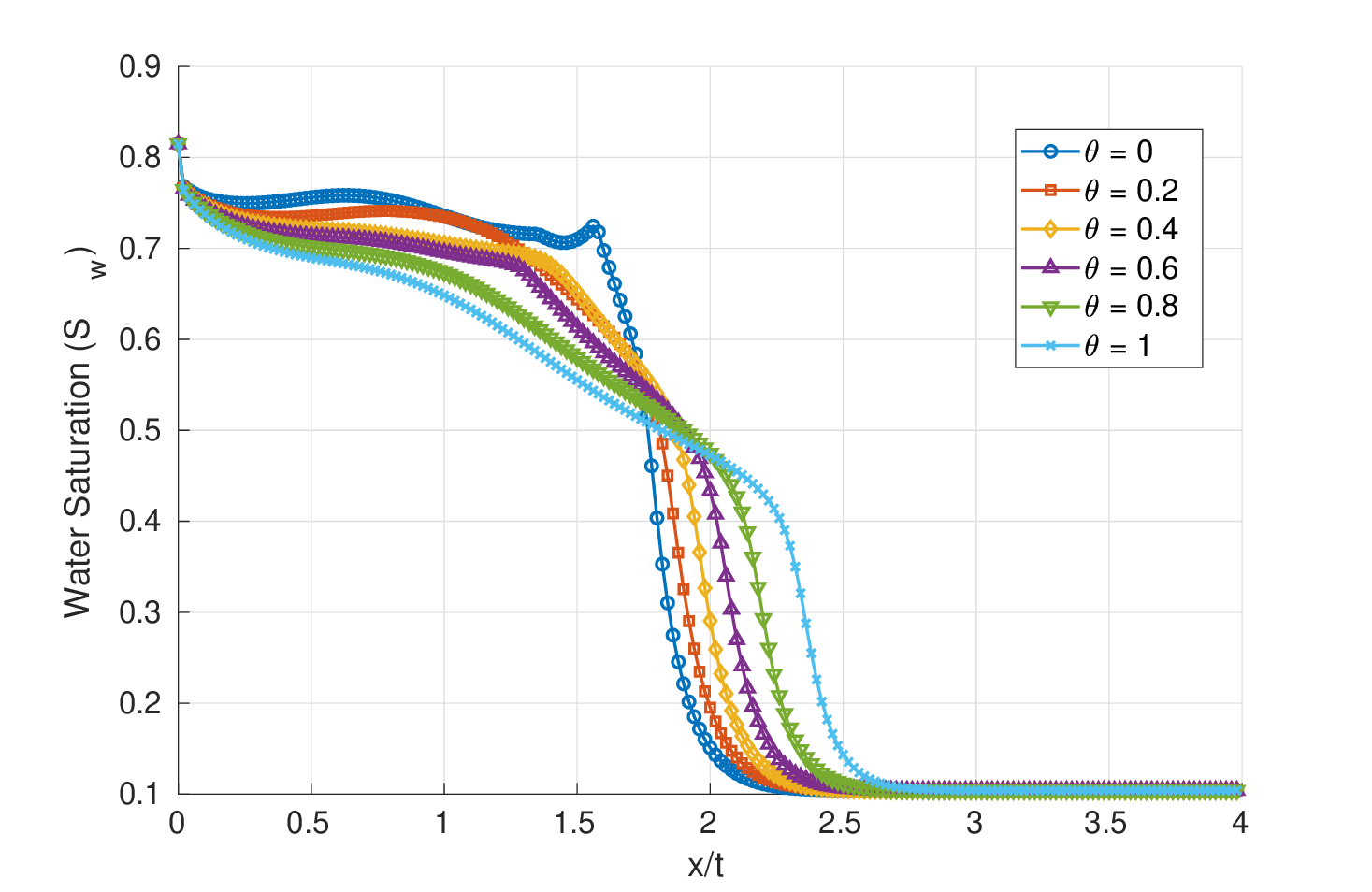}
	 \caption{Water saturation profiles for varying wettability states ($\theta$), showing transition from oil-wet ($\theta = 1$) to water-wet ($\theta = 0$) conditions. Lower $\theta$ values accelerate frontal advance, with 20\% $\theta$ reduction yielding 3\% recovery gain at 2.0 PVI. Dashed vertical lines mark characteristic shock velocities.}
		\label{fig:sw_profiles}
	\end{figure}

	To evaluate the effect of ion-specific interactions, we consider a case where the injected magnesium concentration is increased to 4.2 mol/kgw, compared to an initial concentration of 2.37 mol/kgw. This modification results in an additional 3\% OOIP at 1.5 PVI, consistent with mechanisms described in Section~\ref{sec:Mg_analysis}, where Mg$^{2+}$ disrupts Ca$^{2+}$-carboxylate bonding.
	
	These results emphasize that recovery is sensitive not only to bulk salinity but also to the ionic composition of the brine, particularly in the presence of divalent cations.

Changes in residual oil saturation ($S_{or}$) significantly affect recovery predictions. A 20\% decrease in $S_{or}$ (e.g., from 0.30 to 0.24) leads to approximately 3\% increase in OOIP. This underlines the necessity of accurate experimental determination of endpoint saturations, especially in mixed-wet systems where pore-scale wettability heterogeneity can dominate \cite{mahani2015}.

The sensitivity analysis reveals two primary mechanisms that govern improvements in oil recovery. First, modifying the composition of the injected brine can influence wettability by altering the interpolation parameter $\theta$ through the adjustment of specific ion concentrations, such as [Mg$^{2+}]$ and [SO$_4^{2-}]$, which in turn reduces the TBP. Second, the initial state of the system, captured through the resolution of the Riemann-Goursat problem, determines the configuration of shock and rarefaction waves. These wave dynamics play a crucial role in enhancing the transport of ions, thereby amplifying the effects of ionic contrasts on recovery. These mechanisms highlight the interplay between chemical and dynamic factors in controlling the effectiveness of low-salinity waterflooding.

From a practical standpoint, field-scale implementations should focus on ion-specific optimization—such as adjusting Mg$^{2+}$/SO$_4^{2-}$ ratios—rather than relying solely on bulk salinity reduction. Calibration of the interpolation parameter $\theta$ through core-scale measurements, including two-phase displacement pressure (TBP) and contact angle, is also recommended. Transient simulations of the initial brine replacement process are necessary to account for dynamic wave interactions, contributing to the predictive accuracy of the model. This approach links numerical forecasts with physical mechanisms, providing a structured method for evaluating enhanced oil recovery (EOR) strategies in heterogeneous carbonate reservoirs.

}

\section{Interpreting Salinity in the Context of Equilibrium Geochemistry}
\label{sec:salinity_equilibrium}

The interpretation of salinity effects in low-salinity waterflooding (LSWF) requires distinguishing between the \textit{injected salinity} defined at surface conditions and the \textit{equilibrium salinity} that governs interfacial interactions at reservoir conditions. While injected brines are designed with specific ionic compositions (e.g., \( \text{Cl}^- = 3.93 \, \text{mol/kgw} \), \( \text{Mg}^{2+} = 0.00381 \, \text{mol/kgw} \)), the subsurface system evolves dynamically through geochemical processes that reshape brine chemistry. These include the dissolution of calcite (\( \text{CaCO}_3 \leftrightarrow \text{Ca}^{2+} + \text{CO}_3^{2-} \)), competitive adsorption of potential-determining ions (\( \text{Ca}^{2+} \), \( \text{Mg}^{2+} \), \( \text{SO}_4^{2-} \)) at oil-rock interfaces, and mixing/dilution with connate brine. As demonstrated in Sections~\ref{sec:SO4_analysis}--\ref{sec:Mg_analysis}, these processes collectively determine the effective ionic environment that dictates TBP and wettability alteration.

The apparent contradiction between injected and equilibrium salinity arises from the transient nature of brine-rock interactions. Surface complexation modeling (Section~\ref{sec:TBP_theta_analysis}) reveals that TBP depends not on the injected brine composition alone but on the thermodynamically equilibrated concentrations of key ions at mineral surfaces. For instance, injected \( \text{SO}_4^{2-} \) may become partially sequestered through anhydrite precipitation (\( \text{CaSO}_4 \)), while \( \text{Mg}^{2+} \) competes with \( \text{Ca}^{2+} \) for carboxylate binding sites. This dynamic equilibrium explains why coreflood experiments often report delayed wettability responses despite rapid brine injection \cite{mahani2015}.

In terms of methodology, the model addresses these factors through mass balance equations that integrate ion transport with equilibrium speciation derived from PHREEQC. Initializing the system with connate water chemistry and imposing low-salinity injection as a boundary condition allows the geochemical state to evolve naturally. The resulting equilibrium concentrations of \( \text{Ca}^{2+} \), \( \text{Mg}^{2+} \), and \( \text{SO}_4^{2-} \)—not their injected values—are used to compute TBP via Equation~\eqref{eq:tbp_total}. This approach aligns with experimental observations where wettability alteration correlates with post-equilibrium ionic activities rather than injected brine composition \cite{Zhang2007,austad2015low}.

The reconciliation of injected and equilibrium salinity lies in recognizing that wettability alteration operates at the pore scale, where nanoscale surface reactions override bulk fluid properties. Field-scale implementations must therefore prioritize ion-specific optimization (e.g., \( \text{Mg}^{2+} / \text{SO}_4^{2-} \) ratios) over bulk salinity reduction, as demonstrated by the 12–15\% recovery gains in Ghawar carbonates \cite{Qiao2016}. By associating TBP with the equilibrated ionic environment, the model addresses the salinity paradox, offering a framework aligned with experimental observations and theoretical principles \cite{jerauld2008modeling}.  

This research offers a more refined analysis of the relationship between injected and equilibrium salinity by explicitly considering the dynamic interplay of geochemical reactions and multiphase flow. This allows for a more accurate prediction of optimal injection strategies, bridging the gap between pore-scale mechanisms and field-scale implementation. 
This shift from fixed brine design to dynamic geochemical equilibrium introduces a methodological adjustment for modeling wettability-driven recovery processes.

\section{Conclusion}
\label{con}

This study demonstrates that wettability alteration induced by low-salinity water flooding (LSWF) in carbonates is governed by specific ionic interactions rather than generalized salinity reduction. The TBP, quantified through surface complexation modeling, emerges as a potential indicator of oil–rock adhesion, showing critical sensitivity to Mg$^{2+}$ and SO$_4^{2-}$ concentrations. Numerical experiments indicate that acidic crude oils (TAN higher than 1 mg KOH/g) show higher TBP values compared to sweet crudes, which could be attributed to calcium–carboxylate interactions on calcite surfaces.
Higher pH levels (above 7.5) correlate with increased deprotonation of carboxylic groups, contributing to more hydrophilic conditions.

The synergy between Mg$^{2+}$ (50–200 mmol/kgw) and SO$_4^{2-}$ (higher than 500 mmol/kgw) reduces adhesion by displacing Ca$^{2+}$ from surface sites, highlighting that specific ionic optimization outperforms strategies solely based on salinity dilution. Coupled numerical simulations with PHREEQC and COMSOL accurately reproduce saturation and recovery profiles reported experimentally, validating the integrated geochemical–compositional framework. This approach enables the prediction of up to 14.7\% recovery gains through targeted adjustments in injected water chemistry.

The results underscore the need to design injection strategies that prioritize optimal Mg$^{2+}$/Ca$^{2+}$ and SO$_4^{2-}$/Cl$^{-}$ ratios, particularly in reservoirs with acidic crudes. The predictive capabilities of this framework---rooted in TBP-driven geochemical modeling coupled with multiphase flow simulations---enable efficient screening of LSWF candidates and the design of injection scenarios that maximize oil recovery (e.g., targeting 50--200~mmol/kgw Mg\textsuperscript{2+} and higher than 500~mmol/kgw SO\textsubscript{4}\textsuperscript{2-}) while mitigating risks of unfavorable wettability shifts, such as overdilution-induced calcite dissolution or sulfate scaling. By translating pore-scale ionic interactions into field-relevant metrics, this approach may mitigate operational uncertainty in carbonate reservoir management.
 Future work should explore time-scale effects on surface–fluid equilibria and validate the model in heterogeneous systems with non-equilibrium multiphase flow, broadening its applicability to realistic field conditions.

\begin{sloppypar}
	
	\section*{Acknowledgments}
	Special thanks are extended to Ali A. Eftekhari for his invaluable contribution in reviewing the calculations conducted using the PHREEQC program. The authors express their gratitude to Sergio Pilotto for his unwavering support throughout this research endeavor. Furthermore, the authors acknowledge the generous funding provided by CAPES under grant numbers 88881.156518/2017-01 and 88887.156517/2017-00, as well as CNPq under grants 405366/2021-3 and 306566/2019-2. Additional support from FAPERJ under grants E-26/210.738/2014, E-26/202.764/2017, and E-26/201.159/2021 is also gratefully acknowledged. This study was partially funded by FAPERJ – Carlos Chagas Filho Foundation for Research Support of the State of Rio de Janeiro, Process SEI-260003/006147/2024.
	
\end{sloppypar}

\section{Declarations}
There are no conflicts of competing interests.

\section*{Data Availability}  
The numerical framework and results supporting this study were developed using COMSOL Multiphysics® (License No. 9202103).
The  scripts for PHREEQC and COMSOL are available in the GitHub repository:  
\url{https://github.com/Amaurycruz/wettability.git}
\appendix

\section{Appendix: Coefficients.}
\label{app1a}
Letting \(x\) represent pH (\(-\log_{10}[\text{H}^+]\)), 
\(y\) represents the molar fraction of chloride (\([\text{Cl}^-]\), mol/kgw), 
and \(z\) represents the molar fraction of magnesium  (\([\text{Mg}^{2+}]\), mol/kgw), 
the coefficients are expressed as follows:

We can express the coefficients for the case of type 2 oil as follows.
\begin{align}
	\text{gauss}(x)& = \exp(-x^2) \nonumber
	\\
	\rho_{w1} &= 0.01\cdot (0.06448 + 0.3039 \exp(-0.005546 x^3 z) \nonumber \\
	&- \sin(z) \sin(0.008639 x^2) \cos(z + 0.8679 x) \exp(-0.005546 x^3 z) \nonumber \\
	&- 0.01526 y) \label{coefeq1}
	\\
	\rho_{o1}&=0.01 (0.7981 + 0.1363 \text{gauss}(z + 0.1171 x) \nonumber \\
	&- 0.7981 \text{gauss}(9.645 \text{gauss}(0.2577 x + 0.1349 y)) \nonumber \\
	&- 0.1558 \text{gauss}(z + 0.1171 x) \text{gauss}(9.645 \text{gauss}(0.2577 x + 0.1349 y))^2) 
	\\
	\rho_{r1} &=0.0000001 (4.606 + 0.0343y - 3.094/\tanh(1.418 + z) \nonumber \\
	&- 1.236\exp(-(1.236x + 0.4345z - 0.001579x/z - 2.789 - 0.08079x^2)^2)) \\
	\rho_{o2} &=1.825 + 0.001116x + 0.01245/x + 0.0006694xy \nonumber \\
	&- 0.001236y - 6.613 \times 10^{-5}x^2y + 0.00000000001z 
\end{align}
\begin{align}
	&\rho_{w3} =0.37 y \\
	\rho_{w4} &=0.01 (1.694z^{0.1511} + 0.03021y^2\text{gauss}(3.087 + 0.4009z - 8.76/x) \nonumber \\
	&- 0.1311z - 1.57z^{0.07416}\text{gauss}(3.087 + 0.4009z - 8.76/x)) \\
	\rho_{o4} &=0.004749 + 3.376 \times 10^{-7}y\exp(x) \nonumber \\
	&+ 0.03217\text{gauss}(0.001988\exp(x) + 0.003211z\exp(x + 0.2729y)) \nonumber \\
	&- 0.001043y - 1.609 \times 10^{-6}\exp(x) 
	\\
	\rho_{r4}&=0.000001(0.008068y - 0.0004481x/z \nonumber \\
	&+ 0.9067\text{gauss}\left(\frac{5.256}{2.514x + 0.3922x^2z - 5.256}\right) \nonumber \\
	&- 0.372 - 0.005255x) 
\end{align}
\begin{align}
	\rho_{w5} &=0.37 z
	\\
	\rho_{o5} &= 0.0000001 (0.302 + \frac{1.47}{x} + 0.001829xy^2 - 0.05994z - 0.1314y \nonumber \\
	&- 0.977\text{gauss}(0.1798y + 0.05398x^2)) 
	\\
	\rho_{r5} &=0.00001 ( 0.01103y^2\text{gauss}\left(\frac{48.79}{\exp(0.8058x)}\right)  \nonumber \\
	&+ 0.6058\text{gauss}\left(\frac{0.005821}{z} + \frac{30.78}{x^2 + 0.1733z\exp(0.8058x) - 18.71}\right) ) 
	\label{coefeq21e}
\end{align}
The coefficients presented in \eqref{coefeq1}-\eqref{coefeq21e} serve as inputs for solving the system of conservation laws described by equations \eqref{eq:massbalance1a}-\eqref{eq:massbalance4a}.
\clearpage
\section{Appendix. Chemical Species}
\label{ap:A}
\begin{table}[h!]
	\centering
	\caption{Parameters of the selected chemical reactions.}
	\label{tab:reactions}
	\begin{tabular}{clllc}
		\toprule
		\# & \textbf{Reaction} & \textbf{log K} (100$^\circ$C) & $\Delta H$ (kJ/mol) & \textbf{Reference} \\
		\midrule
		\multicolumn{5}{l}{\textit{Oil and Calcite surface reactions}} \\
		\midrule
		1 & $\text{oil}_{s,\text{NH}^+} \rightleftharpoons \text{oil}_{s,\text{N}} + \text{H}^+$ & -3.61 & 34 & \cite{shock} \\
		2 & $\text{oil}_{w,\text{COOH}} \rightleftharpoons \text{oil}_{w,\text{COO}^-} + \text{H}^+$ & -3.03 & 28 & \cite{shock} \\
		3 & $\text{oil}_{w,\text{COOH}} + \text{Ca}^{2+} \rightleftharpoons \text{oil}_{w,\text{COOCa}^+} + \text{H}^+$ & -3.72 & 1.2 & \cite{shock} \\
		4 & $\text{oil}_{w,\text{COOH}} + \text{Mg}^{2+} \rightleftharpoons \text{oil}_{w,\text{COOMg}^+} + \text{H}^+$ & -3.92 & 1.2 & \cite{shock} \\
		5 & $\text{oil}_{s,\text{NH}^+} + \text{SO}_4^{2-} \rightleftharpoons \text{oil}_{s,\text{NH}_2\text{SO}_4^-}$ & -3.16 & -15 & \cite{shock} \\
		6 & $\text{Cal}_{s,\text{OH}} + \text{H}^+ \rightleftharpoons \text{Cal}_{s,\text{OH}_2^+}$ & 6.75 & -77.5 & \cite{sverjensky} \\
		7 & $\text{Cal}_{s,\text{OH}} + \text{HCO}_3^- \rightleftharpoons \text{Cal}_{s,\text{CO}_3^-} + \text{H}_2\text{O}$ & 11.16 & -61.6 & \cite{sverjensky} \\
		8 & $\text{Cal}_{w,\text{CO}_3\text{H}} \rightleftharpoons \text{Cal}_{w,\text{CO}_3^-} + \text{H}^+$ & -4.52 & 8.3 & \cite{sverjensky} \\
		9 & $\text{Cal}_{w,\text{CO}_3\text{H}} + \text{Ca}^{2+} \rightleftharpoons \text{Cal}_{w,\text{CO}_3\text{Ca}^+} + \text{H}^+$ & -2.52 & 1.2 & \cite{sverjensky} \\
		10 & $\text{Cal}_{w,\text{CO}_3\text{H}} + \text{Mg}^{2+} \rightleftharpoons \text{Cal}_{w,\text{CO}_3\text{Mg}^+} + \text{H}^+$ & -1.88 & 4.5 & \cite{sverjensky} \\
		11 & $\text{Cal}_{s,\text{OH}} + \text{SO}_4^{2-} \rightleftharpoons \text{Cal}_{s,\text{SO}_4^-} + \text{OH}^-$ & -7.55 & -22 & \cite{sverjensky} \\
		\midrule
		\multicolumn{5}{l}{\textit{Aqueous reactions}} \\
		\midrule
		12 & $\text{CO}_2(\text{aq}) + \text{H}_2\text{O} \rightleftharpoons \text{HCO}_3^- + \text{H}^+$ & -6.09 & 7.4 & \cite{supcrtbl} \\
		13 & $\text{HCO}_3^- \rightleftharpoons \text{CO}_3^{2-} + \text{H}^+$ & -9.28 & 14.9 & \cite{phreeqc} \\
		14 & $\text{H}_2\text{O} \rightleftharpoons \text{OH}^- + \text{H}^+$ & -12.25 & 55.8 & \cite{crc} \\
		15 & $\text{CaCO}_3(\text{aq}) \rightleftharpoons \text{Ca}^{2+} + \text{CO}_3^{2-}$ & -8.90 & -12.7 & \cite{phreeqc} \\
		16 & $\text{MgCO}_3 \rightleftharpoons \text{Mg}^{2+} + \text{CO}_3^{2-}$ & -8.15 & -9.8 & \cite{phreeqc} \\
		\midrule
		\multicolumn{5}{l}{\textit{Ion pairing reactions}} \\
		\midrule
		17 & $\text{Ca}^{2+} + \text{H}_2\text{O} \rightleftharpoons \text{CaOH}^+ + \text{H}^+$ & -10.66 & 18.3 & \cite{phreeqc}, \cite{stumm1996aquatic} \\
		18 & $\text{CO}_3^{2-} + \text{Ca}^{2+} + \text{H}^+ \rightleftharpoons \text{CaHCO}_3^+$ & 11.50 & -5.6 & \cite{phreeqc} \\
		19 & $\text{Mg}^{2+} + \text{H}_2\text{O} \rightleftharpoons \text{MgOH}^+ + \text{H}^+$ & -10.02 & 21.0 & \cite{phreeqc} \\
		20 & $\text{CO}_3^{2-} + \text{Mg}^{2+} + \text{H}^+ \rightleftharpoons \text{MgHCO}_3^+$ & 10.98 & -6.1 & \cite{phreeqc} \\
		21 & $\text{Na}^+ + \text{CO}_3^{2-} \rightleftharpoons \text{NaCO}_3^-$ & 1.52 & 3.2 & \cite{phreeqc} \\
		22 & $\text{Na}^+ + \text{HCO}_3^- \rightleftharpoons \text{NaHCO}_3$ & -0.41 & -2.4 & \cite{phreeqc} \\
		23 & $\text{Na}_2\text{SO}_4 \rightleftharpoons 2\text{Na}^+ + \text{SO}_4^{2-}$ & 0.70 & 1.8 & \cite{electrolyte} \\
		24 & $\text{Ca}^{2+} + \text{SO}_4^{2-} \rightleftharpoons \text{CaSO}_4(\text{aq})$ & 2.30 & -10.5 & \cite{phreeqc} \\
		25 & $\text{Mg}^{2+} + \text{SO}_4^{2-} \rightleftharpoons \text{MgSO}_4(\text{aq})$ & 2.50 & -9.7 & \cite{phreeqc} \\
		26 & $\text{Na}^+ + \text{SO}_4^{2-} \rightleftharpoons \text{NaSO}_4^-$ & 0.84 & 0.7 & \cite{phreeqc} \\
		27 & $\text{SO}_4^{2-} + \text{H}^+ \rightleftharpoons \text{HSO}_4^-$ & 1.20 & -22.5 & \cite{supcrt} \\
		\bottomrule
	\end{tabular}
\end{table}
\clearpage
\section{Appendix: Normalized TBP}
\label{ap:B}

\begin{table}[htbp] 
	\centering 
	\caption{Normalized TBP values, calculated using the formula $\theta = {\mathrm{(TBP} - 0.02)}/{(10 - 0.02)}$, for varying concentrations of $\mathrm{Mg^{2+}}$ (0.02–6.80 mol/kgw) and $\mathrm{SO_4^{2-}}$ (0.02–10 mol/kgw) under low-salinity( [Cl]$^{-}$ = 0.056 mol/kgw)} 
	\label{tab:theta_cl2000}
	\begin{tabular}{|c|ccccc|} 
		\hline 
		\textbf{Mg (mol/kgw)} & SO$_4^{2-}$ = 0.02 & 0.04 & 0.06 & 0.08 & 0.12 \\ 
		\hline 
		0.02 & 0.92 & 0.85 & 0.45 & 0.01 & 0.00 \\ 
		0.06 & 0.91 & 0.88 & 0.63 & 0.01 & 0.00 \\ 
		0.16 & 0.87 & 0.89 & 0.82 & 0.73 & 0.01 \\ 
		0.36 & 0.81 & 0.84 & 0.87 & 0.90 & 0.94 \\ 
		0.56 & 0.74 & 0.76 & 0.79 & 0.81 & 0.86 \\ 
		0.86 & 0.65 & 0.67 & 0.69 & 0.71 & 0.74 \\ 
		1.06 & 0.61 & 0.62 & 0.64 & 0.65 & 0.68 \\ 
		1.46 & 0.53 & 0.55 & 0.57 & 0.59 & 0.63 \\ 
		2.00 & 0.47 & 0.48 & 0.50 & 0.51 & 0.56 \\ 
		2.50 & 0.38 & 0.38 & 0.39 & 0.40 & 0.41 \\ 
		3.50 & 0.33 & 0.34 & 0.34 & 0.35 & 0.36 \\ 
		5.50 & 0.30 & 0.30 & 0.31 & 0.31 & 0.31 \\ 
		6.80 & 0.29 & 0.29 & 0.29 & 0.30 & 0.30 \\ 
		\hline 
	\end{tabular} 
\end{table}
\begin{table}[htbp]
	\centering 
	\caption{Normalized TBP values, calculated using the formula $\theta = {\mathrm{(TBP} - 0.02)}/{(10 - 0.02)}$, for varying concentrations of $\mathrm{Mg^{2+}}$ (0.02–6.80 mol/kgw) and $\mathrm{SO_4^{2-}}$ (0.02–10 mol/kgw) under high salinity( [Cl]$^{-}$ = 0.395 mol/kgw)} 
	\label{tab:theta_cl14000}
	\begin{tabular}{|c|ccccc|} 
		\hline 
		\textbf{Mg (mol/kg$_\text{w}$)} & SO$_4^{2-}$ = 0.02 & 0.04 & 0.06 & 0.08 & 0.12 \\ 
		\hline 
		0.02 & 0.90 & 0.91 & 0.93 & 0.94 & 0.88 \\ 
		0.06 & 0.89 & 0.91 & 0.92 & 0.94 & 0.93 \\ 
		0.16 & 0.85 & 0.86 & 0.88 & 0.90 & 0.94 \\ 
		0.36 & 0.85 & 0.79 & 0.80 & 0.82 & 0.89 \\ 
		0.56 & 0.85 & 0.75 & 0.77 & 0.78 & 0.83 \\ 
		0.86 & 0.71 & 0.75 & 0.79 & 0.81 & 0.85 \\ 
		1.06 & 0.69 & 0.70 & 0.73 & 0.75 & 0.80 \\ 
		1.46 & 0.61 & 0.63 & 0.64 & 0.67 & 0.72 \\ 
		2.00 & 0.55 & 0.57 & 0.60 & 0.62 & 0.67 \\ 
		2.50 & 0.37 & 0.37 & 0.45 & 0.39 & 0.40 \\ 
		3.50 & 0.32 & 0.32 & 0.33 & 0.33 & 0.34 \\ 
		5.50 & 0.30 & 0.37 & 0.30 & 0.36 & 0.31 \\ 
		6.80 & 0.28 & 0.29 & 0.29 & 0.29 & 0.30 \\ 
		\hline 
	\end{tabular} 
\end{table}

\section{Appendix. Figures}
\label{app1}
\begin{figure}[htbp]
	\centering
	\includegraphics[width=0.45\linewidth]{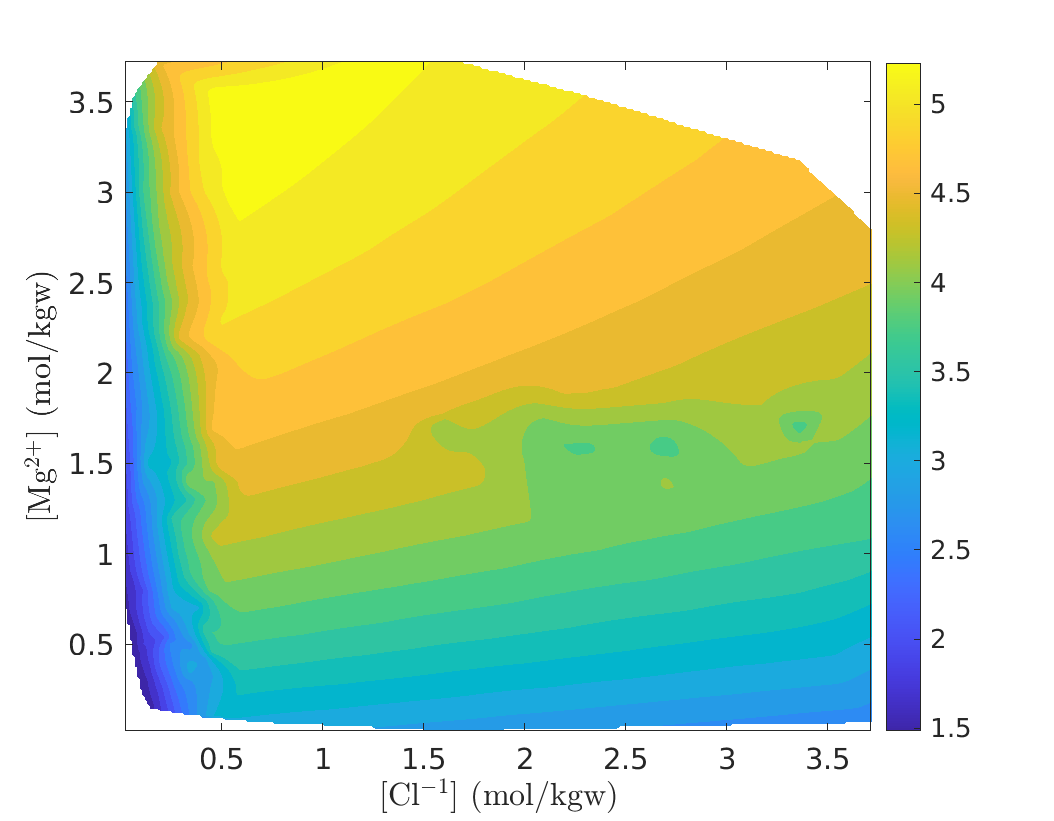}
	\includegraphics[width=0.49\linewidth]{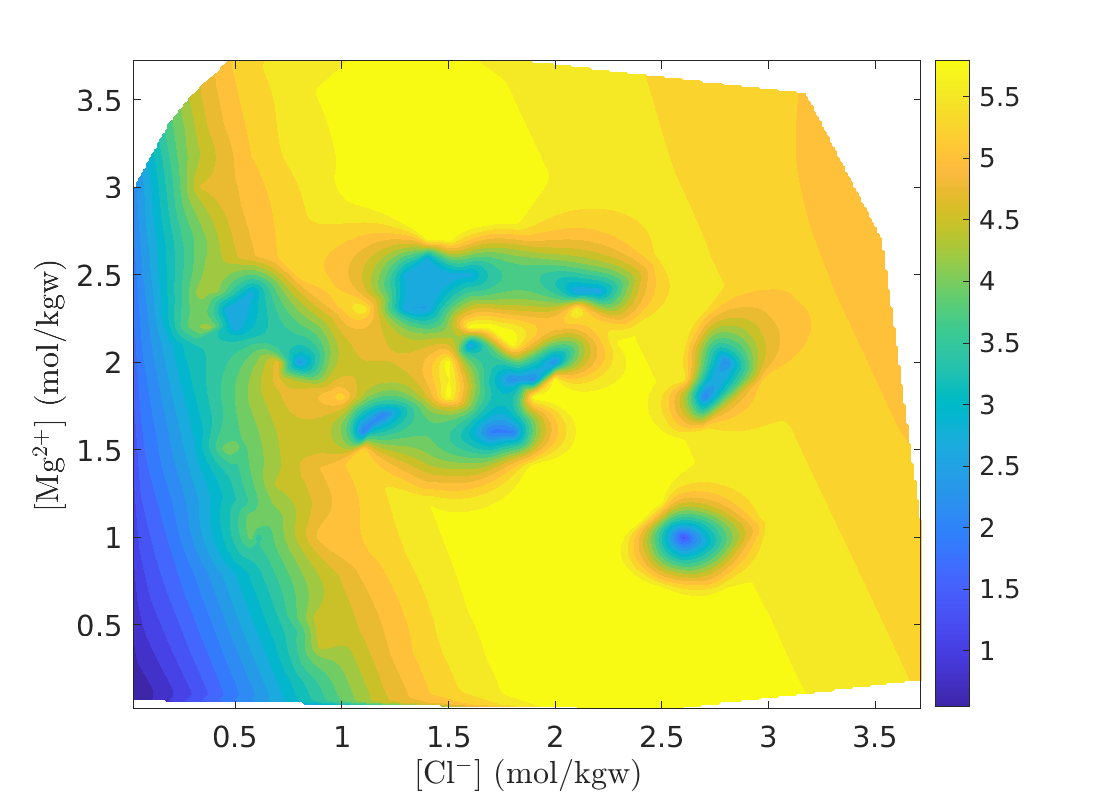}
	\caption{Contour plots of TBP for acid oil and low pH: (a) low [SO$_4^{2-}$] (1 mmol/kg$_\mathrm{w}$), (b) high [SO$_4^{2-}$] (250 mmol/kg$_\mathrm{w}$). Axes: chloride ([Cl$^-$], mol/kg$_\mathrm{w}$) vs. magnesium ([Mg$^{2+}$], mol/kg$_\mathrm{w}$). Color scale indicates TBP values ($\times 10^{-12}$), with lower values (blue) indicating water-wet conditions. TBP values  decrease significantly under high [SO$_4^{2-}$] conditions, particularly at low [Cl$^{-}$] and high [Mg$^{2+}$]. This reduction suggests the synergistic effect of [SO$_4^{2-}$] and [Mg$^{2+}$] in destabilizing oil-rock complexes, which may promoting water-wet behavior.}
	\label{fig:figuras12}
\end{figure}
\begin{figure}[htbp]
	\centering
	\includegraphics[width=0.47\linewidth]{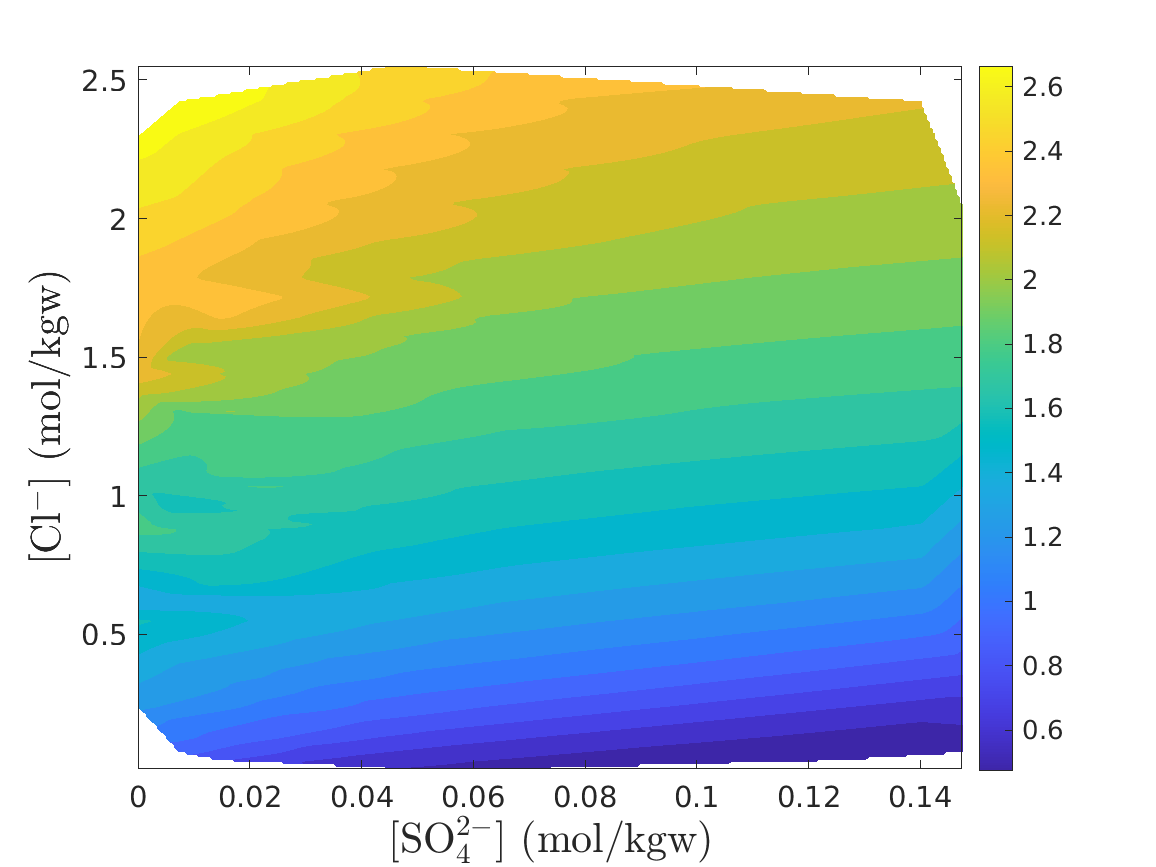}
	\includegraphics[width=0.49\linewidth]{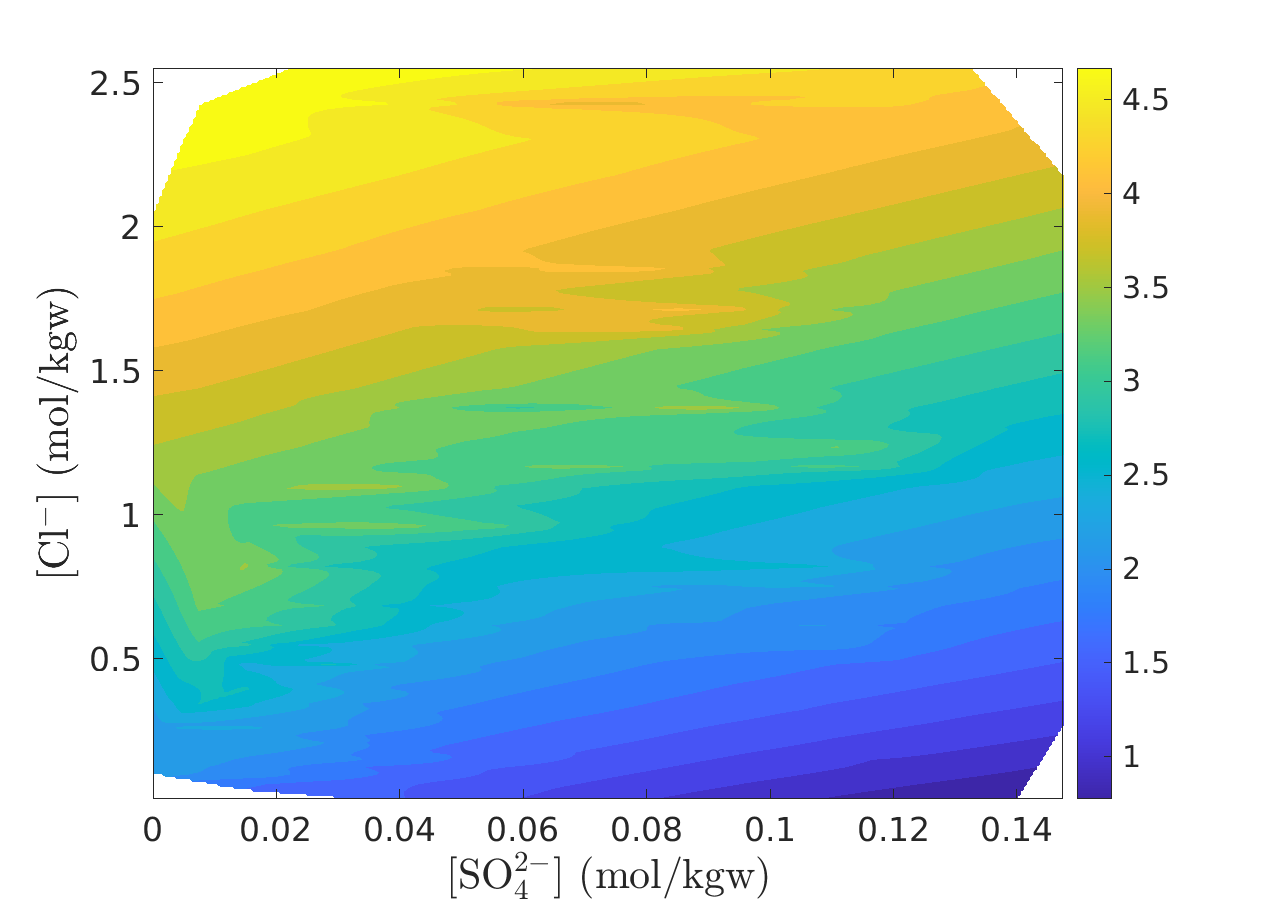}
\caption{Contour plots of TBP for acid oil and low pH: (a) low Mg$^{2+}$ (0.41 mol/kg$_\mathrm{w}$), (b) high Mg$^{2+}$ (10.63 mol/kg$_\mathrm{w}$). Axes: chloride ([Cl$^-$], mol/kg$_\mathrm{w}$) vs. sulfate ([SO$_4^{2-}$], mol/kg$_\mathrm{w}$). TBP values (legend: 0–0.45) decrease with increasing Mg\(^{2+}\), particularly at low Cl\(^{-}\) ($<$0.50 mol/kgw) and high SO\(_4^{2-}\) ($>$0.10 mol/kgw), highlighting the synergistic role of Mg\(^{2+}\) and SO\(_4^{2-}\) in wettability alteration.}
\label{fig:figuras12a}
\end{figure}
\bibliographystyle{unsrt}

	\end{document}